\documentclass{aa}  
\usepackage{placeins}
\usepackage{graphicx}
\usepackage{color,soul}

\usepackage{txfonts}

\usepackage[colorlinks=true,linkcolor=blue,urlcolor=blue,citecolor=blue]{hyperref}

\begin{document}

   \title{The circumstellar environment of the young, low-mass dipper star JH~223}

   \subtitle{Accretion and large-scale magnetic field topology}

   \author{Freitas, T. P.
          \inst{1}
          \and
          Bouvier, J. \inst{2}
          \and
          Zaire, B. \inst{1}
          \and
          Alencar, S. H. P. \inst{1}
          \and
          Sousa, A. P. \inst{2}
          \and
          Rebull, L. \inst{3}
          \and 
          Bayo, A. \inst{4}
          \and
          Frasca, A. \inst{5}
          \and 
          Alonso-Santiago, J. \inst{5}
          \and
          Grankin, K. \inst{6}
          \and
          Contreras Pe\~na, C. \inst{7}
          \and
          Cody, A. M. \inst{8}
          \and
          Hillenbrand, L. A. \inst{9}
          \and
          Carmona, A. \inst{10}}

   \institute{ Universidade Federal de Minas Gerais, Belo Horizonte, MG, 31270-901, Brazil \and Universit\'e Grenoble Alpes, CNRS, IPAG, 38000 Grenoble, France \and Infrared Science Archive (IRSA), IPAC, 1200 E. California Blvd., California Institute of Technology, Pasadena, CA 91125, USA \and European Southern Observatory, Karl-Schwarzschild-Strasse 2, 85748 Garching, Germany \and INAF - Osservatorio Astrofisico di Catania, Via S. Sofia 78, I-95123 Catania, Italy \and Crimean Astrophysical Observatory, 298409, Nauchny, Republic of Crimea \and Department of Physics and Astronomy, Seoul National University, 1 Gwanak-ro, Gwanak-gu, Seoul 08826, Korea \and SETI Institute, 339 N Bernardo Ave, Suite 200, Mountain View, CA 94043, USA \and Department of Astronomy, MC 249-17, California Institute of Technology, Pasadena, CA 91125, USA \and Univ. de Toulouse, CNRS, IRAP, 14 avenue Belin, 31400 Toulouse, France}

   \date{}
 
  \abstract 
   {Studies of magnetospheric accretion and magnetic field topology in T Tauri stars have advanced over the years, but their applications to fully convective, very-low-mass T Tauri stars remain relatively unexplored.}
   {We aim to analyze the circumstellar environment of the very-low-mass dipper-like  star JH 223 by investigating the accretion process and characterizing its large-scale magnetic field topology.}
   {We analyzed the photometric variability of JH 223 using observations from multiple telescopes, including K2, TESS, and LCOGT across different epochs. Additionally, we used Gemini/GRACES spectroscopic and CFHT/SPIRou spectropolarimetric data to investigate the star-disk interaction and to characterize the large-scale stellar magnetic field using Zeeman-Doppler imaging.}
   {JH 223 is a fully convective classical T Tauri star with an age of about 3 Myr and a mass of 0.4 M$_{\odot}$. The large-scale surface magnetic field is predominantly poloidal, with a 250 G dipolar component. The dipole field strength and the mass accretion rate indicate that the disk gas truncation radius is located near the corotation radius ($6 \pm 1$ R$\mathrm{_\star}$). The star-disk interaction, combined with the inclined dipole, generates accretion columns that warp the inner disk. As the star rotates, this warp periodically obscures the stellar surface every 3.31 days, producing the dipper light curves. The same period is also detected in variations of the radial velocity and the longitudinal magnetic field. The accretion columns, traced by strong redshifted absorption components in H$\alpha$ and He I 1083\;nm, are associated with the inner disk warp, as they occur around the same rotational phase. The accretion process in JH 223 is dynamic, transitioning from an unstable to a stable regime over a few weeks, consistent with predictions from magnetohydrodynamic simulations of the star-disk interaction.} 
   {Results from multi-technique observations suggest that the magnetospheric accretion model remains valid for fully convective very-low-mass young stars.}


   \maketitle

\section{Introduction}

Classical T Tauri stars (CTTSs) are young, low-mass stars that accrete material from their surrounding circumstellar disks. These young star-disk systems are observable across a wide range of wavelengths, from X-rays to the millimeter, and typically exhibit strong surface magnetic fields of the order of kilogauss (kG) \citep{Johns-Krull07,Flores22}. The stellar magnetic field plays an important role in these systems, controlling accretion from the disk and outflow through stellar and disk winds \citep{Alencar10, Hartmann16}.

According to magnetohydrodynamic (MHD) simulations \citep{2002ApJ...578..420R,2013MNRAS.431.2673K}, the stellar magnetic field mediates the star-disk interaction, truncating the disk where the magnetic pressure equals the dynamic pressure of the gas. The stellar magnetic field interacts with the inner circumstellar disk, producing accretion columns that channel gas onto the stellar surface. The accreting gas drags the dust accumulated at the base of the accretion funnel, lifting it above the midplane of the disk, and creating a warp at the disk edge \citep{Bodman17}. When the star-disk system is seen at intermediate to high inclinations, the warp hides the stellar photosphere as the system rotates, causing brightness minima in the light curve of T Tauri stars \citep{Mcginnis15}. Stars that exhibit these characteristics are called dippers \citep{Cody14}.

The advantage of observing dipper-like stars is that our line of sight passes through all the relevant structures of the system, including the disk wind, the inner disk region, the accretion funnel, and brightness features on the stellar surface. These systems allow us to investigate whether these structures are correlated as predicted by magnetospheric accretion and ejection models.

\citet{Bouvier99-AATau, Bouvier03} studied the dipper-like star AA Tau and originally proposed that its brightness variability might be caused by an inner disk warp at the base of the accretion column. \citet{Alencar10} analyzed light curves of young stars from the NGC 2264 cluster obtained with the CoRoT satellite and observed that the photometric behavior of AA Tau was not an exception. They found that 28\% of the CTTSs in NGC 2264 exhibited dipper-like variability, suggesting this phenomenon is common. In a subsequent observation campaign of NGC 2264 with CoRoT, \citet{Cody14} significantly expanded the study of young stellar variability in CTTSs. They developed a detailed statistical analysis using metrics of periodicity ($Q$) and symmetry ($M$) to classify the observed light curves. These parameters defined a $Q$-$M$ space suitable for the selection of dipper-like stars. \citet{Mcginnis15} further advanced the understanding of dipper stars by exploring the primary causes of photometric variability in CTTS within NGC 2264. They revealed that dipper-like light curves are predominantly observed in systems with high inner disk inclinations ($i > 59^{\circ}$) with respect to our line of sight and that these light curves can transition from periodic to aperiodic and vice versa over a few years. This variability demonstrates that stars can transition between stable and unstable accretion regimes on these timescales \citep{2016MNRAS.459.2354B}. More recently, \citet{Nagel24} extended the inclination limit above which the dipper phenomenon can be observed to $45^{\circ}$ by considering the dust survival and sublimation along the accretion funnel. These studies have significantly broadened our understanding of the dipper phenomenon in CTTSs, particularly in solar-mass systems. However, the extent to which the magnetospheric accretion model applies to very-low-mass stars still has some open questions. Studies showed that very-low-mass stars and young brown dwarfs can show signals of accretion \citep{Hartmann16,Manara2023}. \citet{Jayawardhana2003} reported that brown dwarfs (i.e., young objects near or below the substellar limit with spectral types between M5 and M8) exhibit detectable accretion rates. \citet{Betti2023} demonstrated a correlation between accretion rate and stellar mass across a broad mass sample, ranging from 2\;M$_\mathrm{\odot}$ to planetary-mass companions, indicating that accretion still occurs in very-low-mass objects. Despite  evidence of accretion in very-low-mass objects, few studies have carried out in-depth analyses of a very-low-mass star using multi-technique, multi-epoch observations, studying the correlations between photometric and spectroscopic variability, the internal structure of the disk, and the role of large-scale magnetic fields in the accretion process. This gap represents one of the main motivations of our work, leading to a more detailed investigation into the applicability of the magnetospheric accretion model for mostly convective stars with masses below 0.5~M$_\mathrm{\odot}$.

To investigate the dipper phenomenon in very-low-mass stars, we focus on JH~223 \citep{1979AJ.....84.1872J}, also known as EPIC 248006676 or 2MASS J04404950+2551191. JH~223 is a dipper-like star ($V = 15.50$) that is a member of the Taurus star formation region \citep{1995ApJS..101..117K}. \citet{1991AJ....101.1050H} classified the source as a weak-line T Tauri star, based on its H$\alpha$ equivalent width $EW_\mathrm{H\alpha}$. However, \citet{2005ApJ...629..881H} later observed a significant near-infrared excess in its spectral energy distribution, attributed to emission from a circumstellar disk, which suggested that JH~223 was an accreting star. \citet{2007ApJ...662..413K} reported the existence of an M6.5 substellar mass companion located at a separation of 2 arcsec ($\simeq$ 300 au). \citet{herczeg2014optical} derived a spectral type M2.8 for the brightest member and no detectable veiling in its low-resolution optical spectrum at 751 nm. From ALMA 1.3 mm observations, where both components of the system have been detected, \citet{2019ApJ...872..158A} derived a primary disk mass of $2 \times 10^{-4}$\;M$_{\odot}$. \citet{Rebull20} reported a rotational period of $3.316 \pm 0.003$\;days\footnote{The error bars were obtained through private communication.} from a K2 light curve and classified the object as a quasiperiodic dipper. \citet{roggero2021dipper} further analyzed the K2 light curve, deriving a period of $3.31 \pm 0.09$\;days and a dip mean amplitude of 0.32 mag. Our initial motivation for selecting JH~223 was based on its K2 dipper light curve, which exhibits a remarkable stability in period and maximum brightness level over many rotations of the system. Its short rotation period has also enabled monitoring several rotational cycles over the course of a ground-based observational campaign. Additionally, the star was observed again years later by the TESS satellite, which revealed significant photometric variability over the years. Taken together, these datasets provide useful constraints on its variability and accretion processes.

The paper is structured as follows. In Sect. \ref{sect:observations}, we present the observations conducted on JH 223 over multiple epochs using several telescopes. These datasets formed the basis for the analyses presented in Sect. \ref{sect:results}, where we detail our results. In Sect. \ref{sect:discussion}, we discuss our results in the context of the star-disk interaction, as predicted by magnetospheric accretion models and simulations. Finally, in Sect. \ref{sect:conclusions} we present our conclusions.

\section{Observations}\label{sect:observations}
In this paper, we analyze multi-epoch observations of JH~223 spanning roughly six years, from 2017 to 2023. The data comprise ground-based photometry, photopolarimetry, high-resolution optical spectroscopy, and near-infrared spectropolarimetry obtained during this campaign, as well as complementary space-based photometry from K2 and TESS. A summary of the observations is available in Tables~\ref{tab:photometric-journal} to \ref{tab:SPIRou_GRACES_journal}, while a detailed description of the instruments, observing sites, and data acquisition is given in Appendix~\ref{apObs}.

\begin{table}
\caption{Journal of photometric observations.}
           
\label{table:phot}      
\centering                          
\begin{tabular}{c c c c }        
\hline\hline                 
Site & HJD & Filters & $N_{meas}$ \\    
& (+2,457,000) & & \\
\hline                        
K2  & 820.6 - 901.1 & - & 3620 \\
CrAO & 1530.8 - 1807.9 &  $V$, $R_J$, $I_J$ & 50 \\
LCOGT & 1773.2 - 1818.3 & $g'$, $r'$, $i'$ & 165 \\
OACT  &  1778.9 - 1905.8 &  $V$, $R_c$, $I_c$ & 186 \\
OACT  &  1778.9 - 1905.8 &  $H\alpha_9$ & 39 \\
OACT &  1778.9 - 1905.8 &  $H\alpha_{18}$ & 78 \\
LT & 1794.2 - 1818.9 & $g'$, $r'$, $i'$ & 36 \\
TESS  & 2474.2 - 2524.5 & - & 31441 \\
TESS & 3208.4 - 3260.0 & - & 30495 \\
\hline                                   
\end{tabular}\label{tab:photometric-journal}
\tablefoot{Col. 1 lists the site or instrument. Col.~2 the Heliocentric Julian Date (HJD) interval of the observations, Col.~3 the photometric filter (when applicable), and Col.~4 the number of measurements. K2 and TESS do not use standard filters, observing instead in broad optical bands spanning approximately 420--900 nm and 600--1000 nm, respectively, covering parts of the standard $BVRI$ and $RIz$ filters.}
\end{table}

\begin{table}
    \caption{Journal of RoboPol photopolarimetric observations.}
    \centering
    \begin{tabular}{ccc}
    \hline\hline
    HJD     & $E$ & Filter\\
    (+2,457,000) & & \\
    \hline
1771.580893 & 0.41 & $R_c$ \\
1771.580891 & 0.41 & SDSS $g$ \\
1787.486404 & 5.21 & $R_c$ \\
\hline
    \end{tabular}
    \label{tab:polshort}
    \tablefoot{Col. 1 lists the heliocentric Julian date, Col.~2 the rotational cycle ($E$) given by the ephemeris in Eq.~\ref{ephemeris}, and Col.~3 the filter used.}
\end{table}

\begin{table}
\caption{Journal of GRACES (G), Keck (K), and SPIRou (S) observations.}
\centering
\begin{tabular}{ccccc}
\hline \hline
Inst. & HJD & $E$ & $t_\mathrm{exp}$ & S/N   \\
        & (+2,457,000) &   & (s) &    \\
\hline
S & 1791.020 & 6.278 & $4 \times 903$   &  96 \\
G & 1794.989 & 7.477 &  1774   & 76  \\
S & 1795.059 & 7.499 & $4 \times 903$  &  82  \\
G & 1795.981 & 7.777 &  1774   & 81 \\
S & 1796.082 & 7.808 & $4 \times 903$  & 68 \\
S & 1796.945 & 8.068 & $4 \times 903$  &  82 \\
S & 1798.117 & 8.422 & $4 \times 903$  &  85  \\
S & 1799.039 & 8.701 & $4 \times 903$  &  78 \\
G & 1800.870 & 9.254 & 1774    & 84 \\
S & 1800.969 &  9.284 & $4 \times 903$   &  89 \\
S & 1801.984 & 9.591 & $4 \times 903$  &  98 \\
K & 1816.795 & 14.065  &  1870   & 110 \\
K & 1817.108 & 14.160 &  1500   & 71 \\
K & 1818.886 & 14.697 &  1870   & 44 \\
G & 1868.882 & 29.802 &  1774   & 73 \\
\hline
\end{tabular}    
  \label{tab:SPIRou_GRACES_journal}
  \tablefoot{Cols. 1 to 4 list the instrument, the Heliocentric Julian Date, the rotational cycle, and the exposure time. For the GRACES and Keck data,  Col.~5 gives the average S/N measured around 750\;nm, while for the SPIRou data, Col.~5 gives the average S/N around 1650\;nm.}
\end{table}

\section{Results}\label{sect:results}

\subsection{Photometry}\label{Photometry}

\begin{figure}
   \centering
   \includegraphics[width=0.75\linewidth]{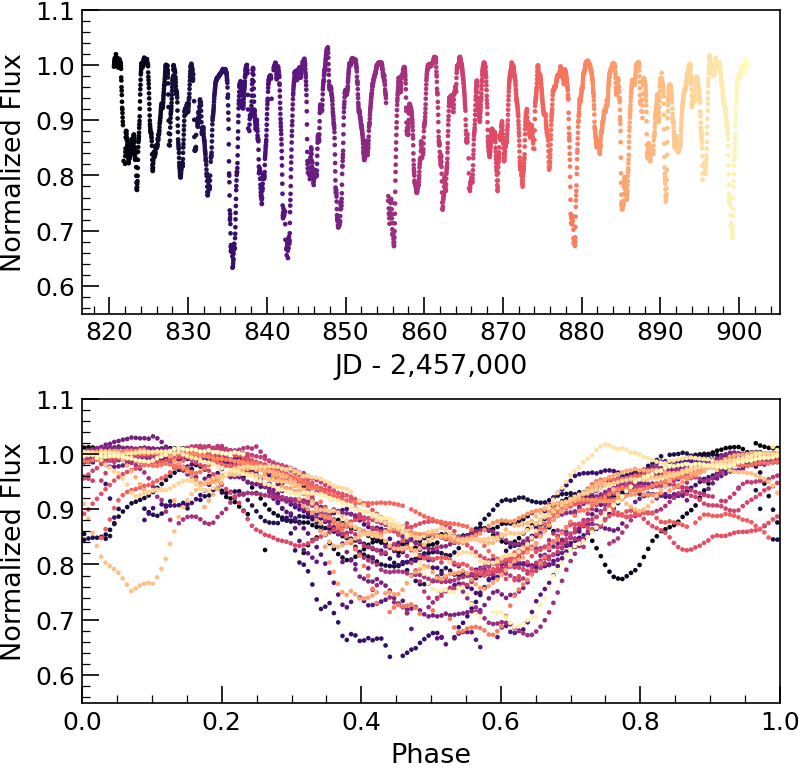}
   \caption{JH~223 K2 light curve shown as a function of Julian date (top) and rotational phase (bottom). The rotational phase was computed using a period of 3.31\;days, with the reference time  $\mathrm{JD}_0 = 2,457,817.63$ chosen such that phase 0.5 corresponds to the photometric minimum. The color coding indicates to the Julian date of each observation.}
              \label{fig:k2lc}
    \end{figure}

The K2 light curve of JH~223 is shown in Fig.~\ref{fig:k2lc} and exhibits a clear dipper behavior, with a constant maximum flux level interrupted by quasiperiodic luminosity dips \citep{Cody14}. A periodogram analysis performed on this light curve returns a period of $3.31 \pm 0.03$ days, where the formal error is measured as the $\sigma$ of the Gaussian fit of the periodogram peak in the frequency space. We modeled the light curve with a periodic kernel as a Gaussian process that yielded $3.32 \pm 0.01$ days. Although the period remains consistent within the error bars, we chose to adopt the more conservative uncertainty derived from the periodogram fit, using a period of $3.31 \pm 0.03$ days in the analyses presented in this study. The light curve folded in phase at this period is also shown in Fig.\;\ref{fig:k2lc}. 

\begin{figure}
   \centering
   \includegraphics[width=\hsize]{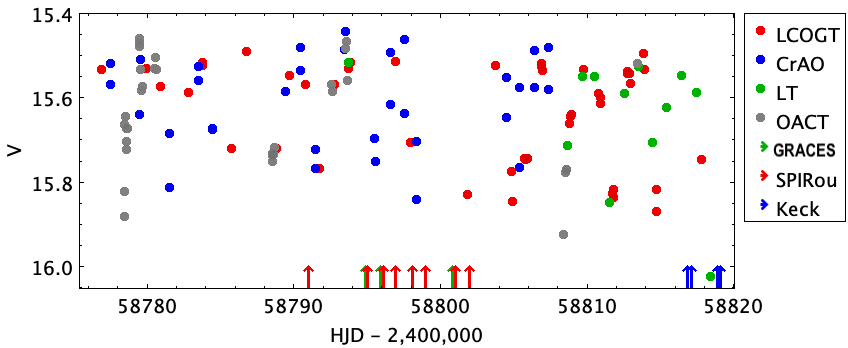}
   \includegraphics[width=\hsize]{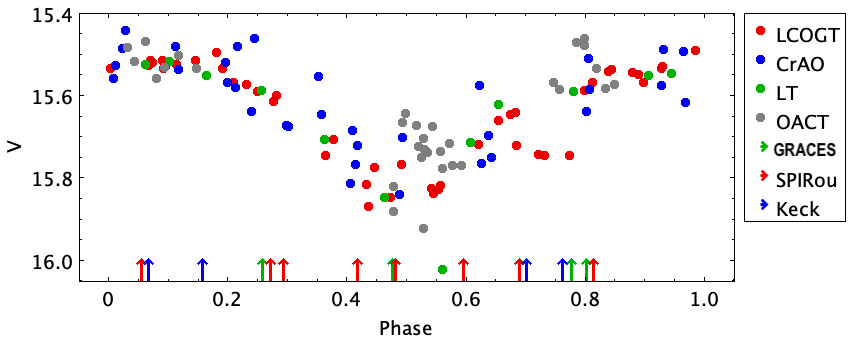}
    \caption{JH~223 $V$-band light curve versus HJD (top) and rotational phase (bottom). The phased light curve uses a period of $3.31$\;days and $\mathrm{JD}_0 = 2{,}458{,}770.2388$ (phase 0.5 at minimum). Arrows mark the epochs of spectroscopic and spectropolarimetric observations. CrAO and OACT data show larger scatter due to higher uncertainties.}
               \label{fig:lcv}
\end{figure}

\begin{figure}
   \centering
      \includegraphics[width=\hsize]{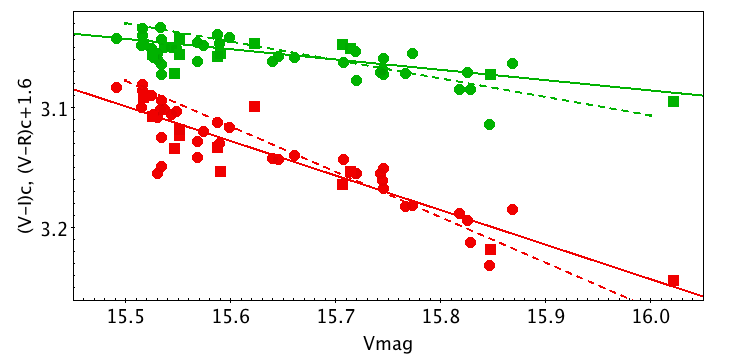}
\caption{CMDs of $(V-R)_c$ vs. $V$ (green) and $(V-I)_c$ vs. $V$ (red), with $(V-R)_c$ shifted by $+1.6$\;mag for clarity. LCOGT (circles) and LT (squares) data are shown. Solid and dashed lines indicate least-squares fits and ISM-extinction slopes, respectively.} 
    \label{fig:colmag}
    \end{figure}

Figure~\ref{fig:lcv} shows the $V$-band light curve of JH~223 obtained during our campaign as a function of the Julian date and folded in phase with the photometric period of 3.31 days, as derived from the K2 observations. The amplitude of the $V$-band light curve is similar to that of the K2 light curve obtained more than 2 years earlier, with a maximum of $m_{V} = 15.49$, and the periodicity has clearly been recovered. The phased light curve exhibits a flat maximum brightness level and relatively deep dips ($\sim$30\%), whose depth and shape vary from one cycle to the next, as also seen in the K2 light curve. This type of light curve is characteristic of dippers and quite different from spot modulation, which usually produces smooth, sinusoidal-like shallower light curves that are stable over a timescale of a few weeks. Therefore, we ascribe the photometric variability of JH~223 to the recurrent occultation of the central star by circumstellar dust located close to corotation. A similar photometric behavior and periodicity can be seen in the $R_c$ and $I_c$ light curves of JH~223. The photometric periodicity and phase appear to have been retained over the full range of observations (i.e., over more than a year) from HJD 58530 (CrAO) to HJD 58905 (OACT), a time span corresponding to 113 rotational periods. The color-magnitude diagrams (CMDs) shown in Fig.~\ref{fig:colmag} indicate that the system becomes redder when it gets fainter. Furthermore, the color slope is shallower than the reddening slope expected from ISM-type grains \citep{cardelli1989relationship}, which indicates that larger grains are responsible for the stellar occultations. For the 2019--2020 observational campaign, which includes contemporaneous ground-based photometry, photopolarimetry, spectroscopy, and spectropolarimetry, we adopted the following ephemeris,
\begin{equation}\label{ephemeris}
HJD(d) = 2{,}458{,}770.2388 + 3.31E,
\end{equation} \label{eq:ephemeris}
\noindent where $E$ is the rotational cycle. For photometric datasets separated by more than one year from this campaign, we adopted different $\mathrm{JD}_0$ values to keep the photometric minima around phase 0.5. This approach is required because the uncertainty in the rotational period accumulates over the years, leading to a loss in terms of the phase coherence.
      
\begin{figure}
   \centering
   \includegraphics[width=0.75\linewidth]{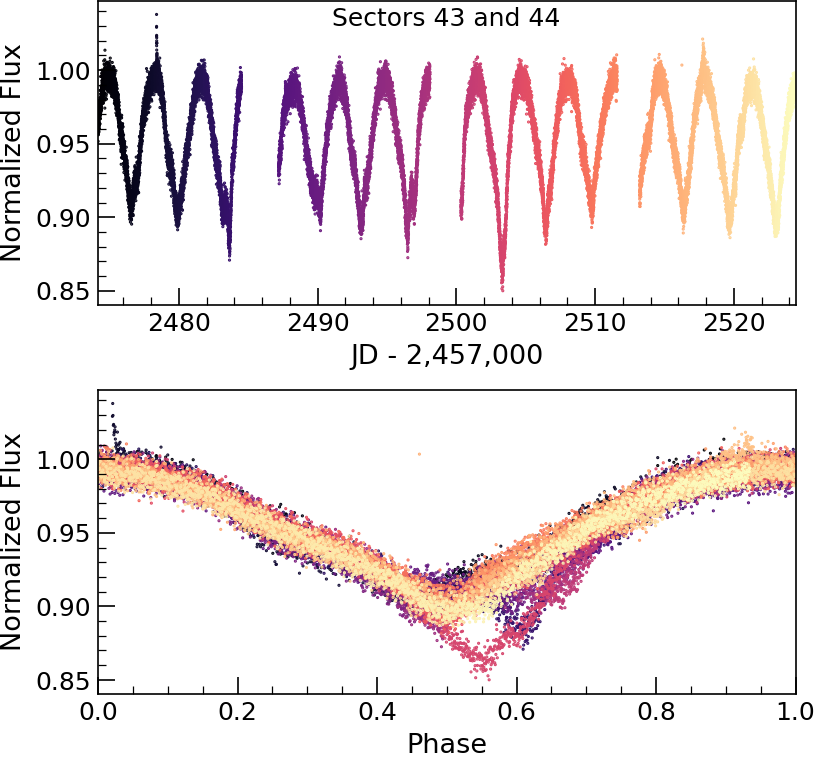}
   \caption{Same as Fig.~\ref{fig:k2lc}, but for the 2021 TESS light curve from sectors 43 and 44. The rotational phase was computed with the 3.31-day period and a reference time of $\mathrm{JD}_0 = 2{,}459{,}471.7$.}
   \label{fig:tess4344}
    \end{figure}

\begin{figure}
   \centering
   \includegraphics[width=0.75\linewidth]{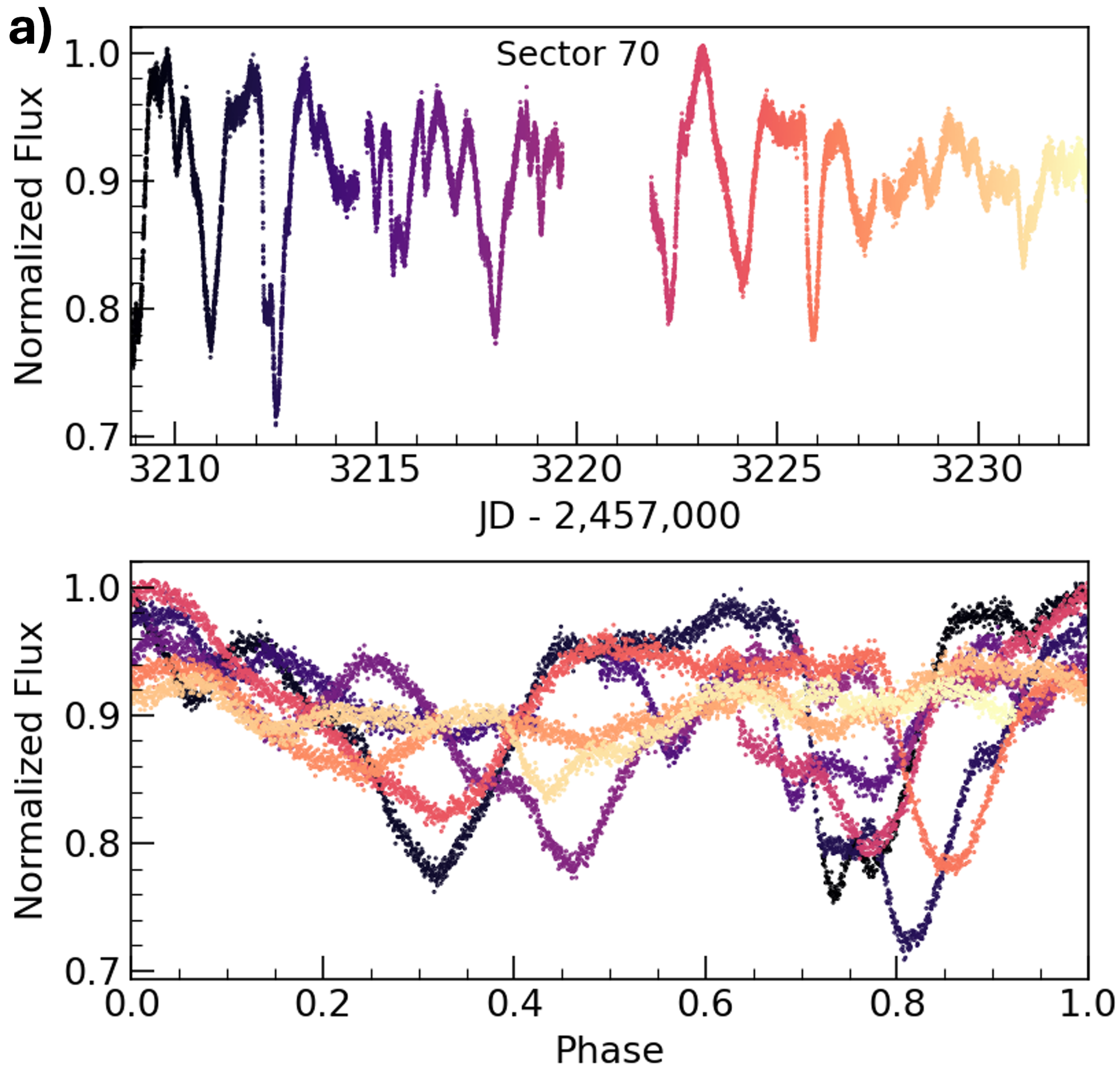}
   \includegraphics[width=0.75\linewidth]{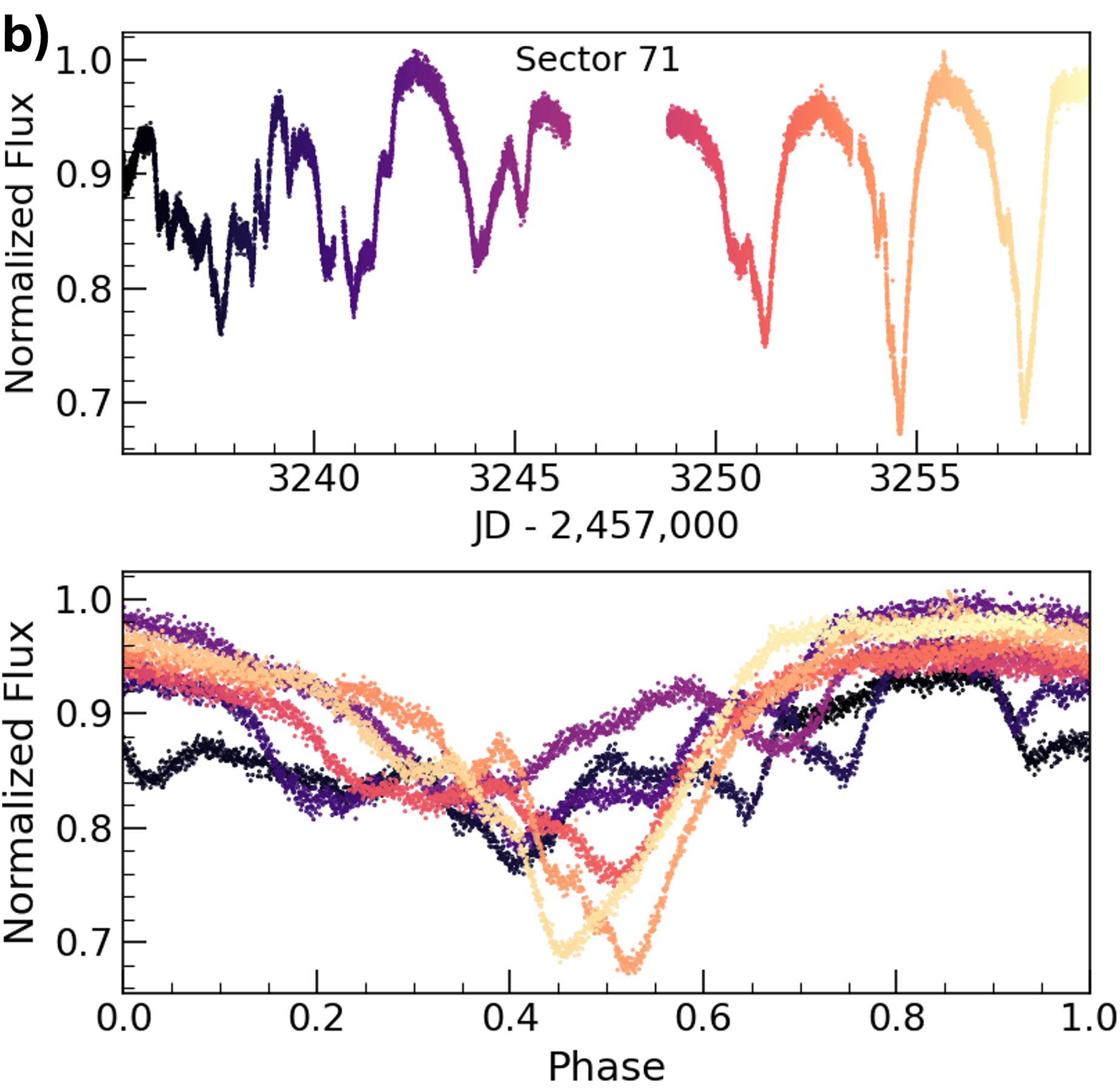}
   \caption{Same as Fig.~\ref{fig:k2lc}, but for the 2023 TESS light curves from sectors 70 (a) and 71 (b). The same period and reference time as in Fig.~\ref{fig:tess4344} were used.}
    \label{fig:tess7071}
    \end{figure}

The TESS satellite observed JH~223 during two epochs, in 2021 and 2023. Figures \ref{fig:tess4344} and \ref{fig:tess7071} show the light curves from these two observation periods. The 2021 light curve (sectors 43 and 44) shows smoother and more regular variability than the K2 and ground-based light curves, and it seems to be dominated by spots rather than dust obscuration, as compared to the other epochs of observation. 

The 2023 light curve presents two distinct behaviors in sectors 70 and 71, as can be seen in Fig. \ref{fig:tess7071}. In sector 70, the variations are complex, and a periodogram analysis suggests a period of 1.66\;days (i.e., half the period derived at other epochs). The latter is gradually recovered in sector 71, whose light curve presents similar patterns to the K2 light curve, with quasiperiodic brightness dips. The periodogram analysis of the TESS light curves returned period values consistent with the period from the K2 light curve: $3.3 \pm 0.1$\;days for the 2021 observations and $3.4 \pm 0.2$\;days for sector 71 of the 2023 observations. We still chose to use the period obtained from the K2 light curve in our analysis, since it is more precise, covering a larger number of rotational cycles, as compared to the TESS light curves. The phase-folded TESS light curves with the K2 adopted period are displayed in the bottom panels of Figs. \ref{fig:tess4344} and \ref{fig:tess7071}.

\subsection{Optical spectroscopy}

\subsubsection{Stellar parameters}\label{stellar_param}

We determined the stellar effective temperature ($T_\mathrm{eff}$), logarithm of surface gravity ($\log {g}$), projected rotational velocity ($v\sin{i}$), and radial velocity ($\upsilon_\mathrm{rad}$). We fit the synthetic spectra to the GRACES data, using the spectral synthesis code \verb|ZEEMAN| \citep{landstreet1988magnetic, folsom2016evolution}. This code computes synthetic spectra based on the MARCS atmosphere model \citep{plez2008marcs, 2008A&A...486..951G} and spectral lines information from the Vienna Atomic Line Database \citep[VALD,][]{2015PhyS...90e4005R}. We assumed solar metallicity. The statistical significance of the fit is evaluated from a $\chi^{2}$ minimization method based on the Levenberg-Marquardt algorithm (LMA).

\begin{figure}[ht]
   \centering
   \includegraphics[width=0.8\linewidth]{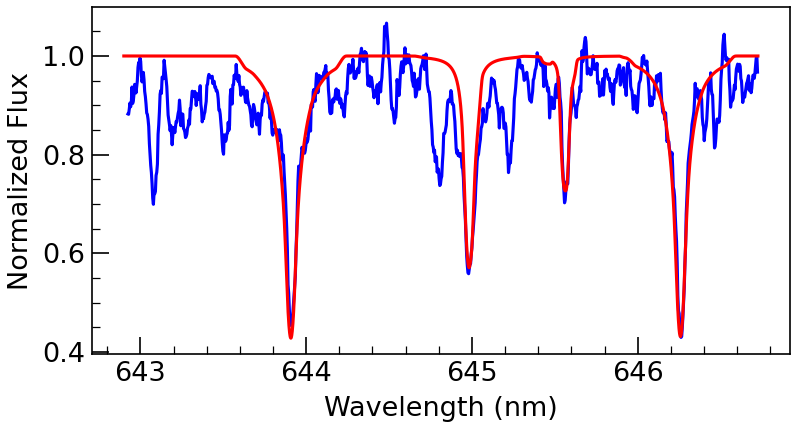}
   \caption{Example of a spectral window used for the stellar parameter fitting. The corresponding synthetic spectrum (red line) is overplotted on the observed GRACES spectrum (blue line).}
              \label{fig:fit_params}
\end{figure}

We chose the spectrum collected on November 13, 2019 for the following analysis, as it presented the highest signal-to-noise ratio (S/N) among our observations. We ran the fitting procedure independently on nine windows within the 640\;nm to 870\;nm wavelength range, excluding the regions with molecular, telluric, and emission lines. Using multiple spectral windows helps to determine the stellar parameters more precisely, as different photospheric lines along the spectrum have different sensitivities to $T_\mathrm{eff}$ and $\log {g}$. We adjusted one parameter at a time, namely, when fitting $T_\mathrm{eff}$, we kept the other parameters fixed. This step was done iteratively for all parameters until we found a converged set of parameters that reproduced the window of interest. Figure~\ref{fig:fit_params} shows one spectral window used for stellar parameter determination overplotted with the corresponding synthetic spectra. We computed the average values obtained from the independent windows, excluding spectral windows that presented values larger than $1 \sigma$ of the mean of each parameter, which provided the following optimal values for the parameters, $T_\mathrm{eff} = 3528 \pm 213$\;K, $\log {g} = 3.9\pm 0.1$, $v\sin{i} = 14.6 \pm 0.9$\;km/s and $\upsilon_\mathrm{rad} = 16.2 \pm 0.4$\;km/s. The radial velocity value listed in this section corresponds to the measurement obtained at a specific rotational phase, associated with the spectrum used for the determination of the stellar parameters. The adopted radial velocity of the star was actually derived from measurements taken at various rotational phases and is presented in Sect. \ref{sec:RV}. The uncertainty associated with each parameter was determined by evaluating the standard deviation of the values obtained from individual spectral windows. The final parameter values are presented in Table \ref{tab:stellar_param}. The stellar parameters ($T_\mathrm{eff}$, $\log {g}$, and $v\sin{i}$) determined with optical data agree within 1 $\sigma$ with values obtained by \cite{Jonsson20} in the near-infrared using APOGEE-2 data.

Using the best-fit parameters obtained for each spectral window from the GRACES spectra, we computed theoretical models for the Keck spectra, disregarding two windows that were unavailable as they fell within the inter-order gaps of the spectra. Keeping all parameters fixed except for the radial velocity, which was determined individually for each observation, we found that the resulting fits are consistent with the observed spectra. The radial velocity values for each observation were calculated by averaging the results obtained from each spectral region. The final mean radial velocity values obtained for each observation were $15.38 \pm 0.24$\;km/s, $16.17 \pm 0.27$\;km/s, and $14.74 \pm 0.33$\;km/s, respectively. The uncertainties were assessed based on the variation in \(\chi^2\) around the minimum value identified using the \verb|ZEEMAN| code for different radial velocity values within each spectral window. The final error bar was determined as the average of the individual uncertainties.

We also computed the luminosity from the relationship between bolometric magnitude and stellar luminosity. We used the maximum magnitude on the $V$ band obtained from the LCOGT observations, $m_{V} = 15.49$ (see Sect. \ref{Photometry}), as the stellar magnitude. We further considered the extinction $A_{V} = 1.2 \pm 0.2$ from \citet {herczeg2014optical}, the distance $d = 143 \pm 3$\;pc from Gaia EDR3 \citep{brown2021gaia}, and we derived the bolometric correction $BC_{V} = -1.91$, interpolating for intermediate spectral types from the tables of \citet{pecaut2013intrinsic}. We then obtained the bolometric magnitude $M_\mathrm{bol} = 6.6 \pm 0.2$, resulting in a stellar luminosity $L_\mathrm{\star} = 0.18 \pm 0.03$\;L$_\mathrm{\odot}$. The luminosity we derived agrees with previous values reported in the literature of $0.17 \pm 0.01$\;L$_{\odot}$  \citep{herczeg2014optical} and $0.21 \pm 0.03$\;L$_{\odot}$ \citep{roggero2021dipper}. 

Taking the values of  $T_\mathrm{eff}$ and $L_\mathrm{\star}$ determined in this paper (Table~\ref{tab:stellar_param}), we obtained the stellar radius $R_\mathrm{\star} = 1.1 \pm 0.2$\;R$_\mathrm{\odot}$ from the Stefan-Boltzmann law. This radius, combined with $v\sin{i}$ and the stellar rotational period, was used to derive the inclination of the star with respect to our line of sight, $i = 57^{\circ} \pm 14^{\circ}$. This inclination, within the error bars, is consistent both with the inner disk warp model discussed by \citet{Mcginnis15}, which requires an inclination greater than $59^{\circ}$ for the dipper phenomenon to be observed, and with the results obtained by \citet{Nagel24}, which predict that the dipper phenomenon can be observed in light curves at inclinations larger than $45^{\circ}$, taking into account dust sublimation along the accretion funnel.

Using pre-main sequence (PMS) evolutionary tracks from \citet{baraffe2015new}, we derive the stellar mass of $0.4 \pm 0.2$\;M$_{\odot}$ and an age of $\sim 3$~Myr from the location of JH~223 in the Hertzsprung-Russell diagram (HRD; see Fig.~\ref{fig:confusogram}). The evolutionary model also indicates that JH~223 is fully convective. Table \ref{tab:stellar_param} summarizes the main stellar parameters of JH~223.

\begin{table}[htbp]
\caption{Stellar parameters of JH~223 determined in this work.}
\centering 

\begin{tabular}{lll}
\hline
\hline
Parameters & Values & Reference \\ 
 & & \\ \hline
Age & $\sim 3$ Myr  & Sec.\;\ref{stellar_param} \\ 
$T_\mathrm{eff}$  & $3528 \pm 213$\;K & Sec.\;\ref{stellar_param}  \\ 
$\log {g}$ & $3.9 \pm 0.1$ & Sec.\;\ref{stellar_param}  \\
$v\sin {i}$  & $14.6 \pm 0.9$\;km/s  & Sec.\;\ref{stellar_param}    \\ 
$\upsilon_\mathrm{rad}$ & $15.3\pm 0.6$\;km/s  & Sec.\;\ref{sec:RV} \\ 
$P$  & $3.31 \pm 0.03$\;d & Sec.\;\ref{Photometry} \\ 
$L_\mathrm{\star}$ & $0.18 \pm 0.03$\;L$_\mathrm{\odot}$ & Sec.\;\ref{stellar_param} \\ 
$M_\mathrm{\star}$ & $0.4 \pm 0.2$\;M$_{\odot}$  & Sec.\;\ref{stellar_param}\\ 
$R_\mathrm{\star}$ & $1.1 \pm 0.2$\;R$_\mathrm{\odot}$  & Sec.\;\ref{stellar_param}\\
$i$  & $57\degr \pm 14\degr$  & Sec.\;\ref{stellar_param}\\ 
$\dot{M}_{\mathrm{acc}}$ &  $(7 \pm 1)\times10^{-11}$\;M$_{\odot}$\;yr$^{-1}$ & Sec.\;\ref{sec: Mass accretion rate}\\
$r_\mathrm{cor}$  &  $6 \pm 1$\;R$_\mathrm{\star}$ & Sec.\;\ref{sect:discussion} \\  
$r_\mathrm{m}$   &   $6 \pm 1$\;R$_\mathrm{\star}$ &  Sec.\;\ref{sect:discussion} \\
\hline
\end{tabular}

  \label{tab:stellar_param}
\end{table}

\subsubsection{Veiling}\label{sec:optical_veiling}

The spectra of accreting T Tauri stars exhibit veiled photospheric lines compared to non-accreting systems. From the ultraviolet to the optical, veiling originates from the continuum emission of the accretion shock at the stellar surface \citep{Hartmann16}.

To measure the optical veiling of JH~223, we selected four regions in the GRACES spectra between $534.5$\;nm and $758.0$\;nm. These regions were chosen for their high S/N and significant number of photospheric lines. We used the ESPaDOnS spectra of the weak-line T Tauri star TWA 7 as a template. TWA 7 has an age of 17 Myr, a mass of $0.62 \pm 0.03$\;M$_{\odot}$ and $v\sin {i} = 4.5 \pm 0.2$\;km/s \citep{nicholson2021surface}. The spectral type of this star varies in the literature: according to \citet{1999ApJ...512L..63W}, TWA 7 has a spectral type M1, while \citet{2013A&A...551A.107M} quote an M2 spectral type and \citet{herczeg2014optical} list an M3.2 spectral type. We broadened the normalized spectrum of TWA 7 according to the projected rotational velocity of JH~223 ($v\sin{i} = 14.6 \pm 0.9$\;km/s) and corrected the shift due to the radial velocity difference between the two stars. Finally, we calculated the veiling ($r$), defined as the ratio of the continuum excess flux to the stellar flux ($r = F_\mathrm{Excess}/F_\mathrm{Phot}$), through a $\chi^2$ minimization procedure to match the spectra of JH~223 with TWA 7 \citep{2023arXiv230102450S}.

Table \ref{tab:GRACES_veiling} shows the mean veiling values of all the spectral regions between $534.5$\;nm to $758.0$\;nm, computed for each night of observation. The error was determined as the standard deviation of the veiling measurements obtained from the selected spectral regions. We derived the average veiling of $0.12 \pm 0.05$, where the error was calculated using the standard deviation of all measurements. We conclude that the star presents low optical veiling, which is constant in our observations within the errors, indicating a low mass accretion rate (see Sect. \ref{sec: Mass accretion rate}). Our low optical veiling values are consistent with those derived using the \verb|ZEEMAN| code in Section~\ref{stellar_param}, both indicating low veiling in the 600--750\;nm range. These results are also in agreement with those reported by \citet{herczeg2014optical}, who found no veiling at optical wavelengths.

\begin{table}
\caption{Optical veiling of JH 223 obtained from GRACES and Keck data.} 
\centering
\begin{tabular}{cccccc}
\hline \hline
HJD & $E$ & Veiling \\
(+2,457,000) &   &   \\
\hline
1794.989 &  7.477 &  $0.18 \pm 0.07$   \\
1795.981 & 7.777 & $0.15 \pm 0.03$ \\
1800.870 & 9.254 & $0.18 \pm 0.07$  \\
1816.795 &  14.065 &  $0.05 \pm 0.04$   \\
1817.108 & 14.160 & $0.06 \pm 0.01$ \\
1818.886 & 14.697 & $0.09 \pm 0.01$  \\
1868.882 & 29.802 &  $0.16 \pm 0.02$   \\
\hline
\end{tabular}
  \label{tab:GRACES_veiling}
  \tablefoot{Cols. 1 to 3 represent the Heliocentric Julian Date of mid-observation, the rotational cycle ($E$), and the mean veiling of the selected spectral regions from $534.5$\;nm to $758.0$\;nm.}
\end{table}

\subsubsection{Circumstellar lines}

\begin{figure}[ht]
   \centering
   \includegraphics[width=0.8\linewidth]{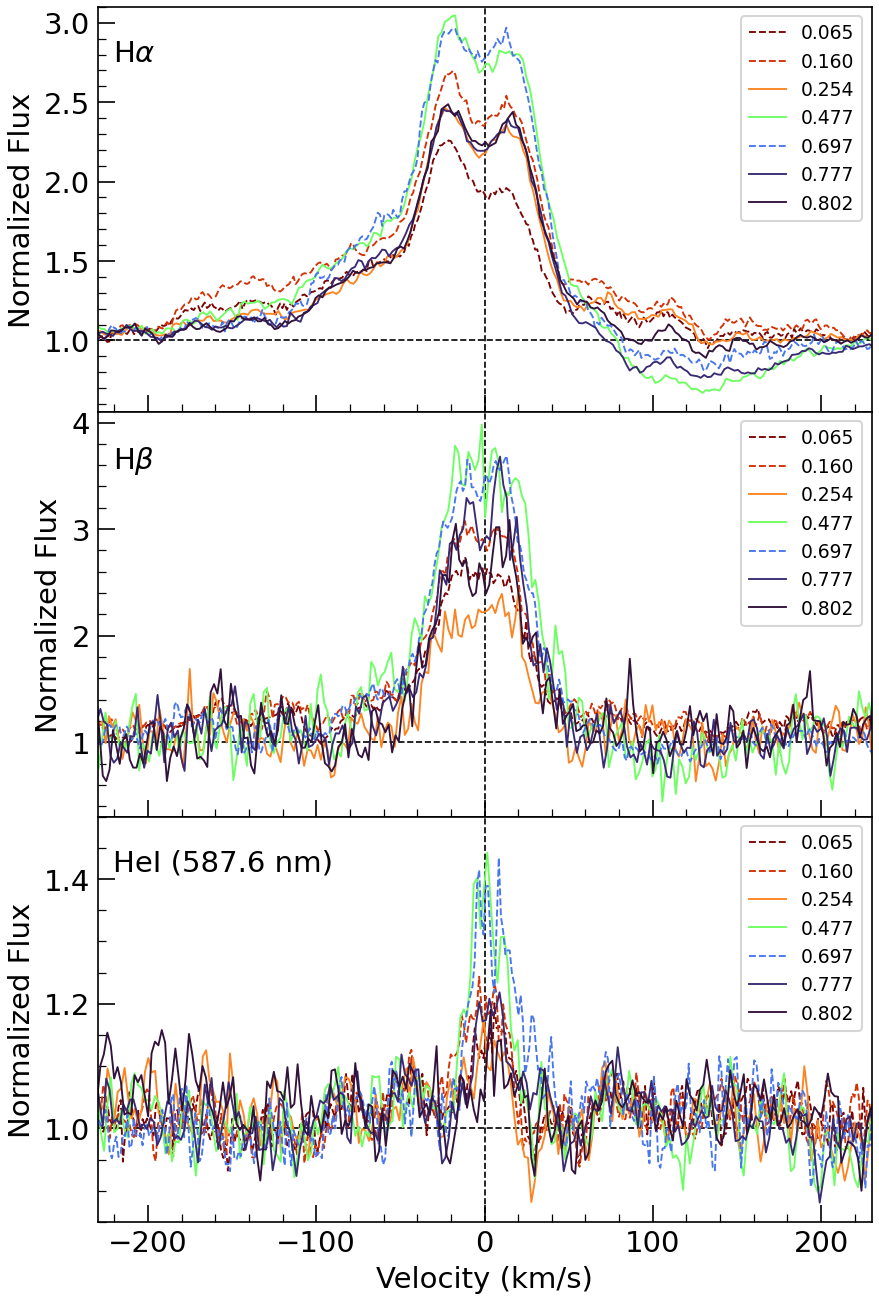}
    \caption{$\mathrm{H{\alpha}}$ (top), $\mathrm{H{\beta}}$ (middle), and He I 587.6 nm (bottom) emission lines used to derive mass accretion rates. Labels indicate the rotational phase. GRACES and Keck data are shown as solid and dashed lines.}
              \label{fig:halphasHBetaGRACESkeck}
\end{figure}

We analyzed the $\mathrm{H{\alpha}}$, $\mathrm{H{\beta}}$, and He I 587.6\;nm lines in the GRACES and Keck spectra of JH 223. The $\mathrm{H{\alpha}}$ and $\mathrm{H{\beta}}$ lines are formed along the accretion funnel, while He I 587.6\;nm originates near the accretion shock, making them good tracers of accretion in the optical domain. We removed the photospheric contribution from the JH~223 spectra to obtain only the circumstellar component by subtracting from the JH~223 spectra a veiled and rotationally broadened mean spectrum of TWA 7 (see Sect. \ref{veiling}). Figure~\ref{fig:halphasHBetaGRACESkeck} shows the circumstellar profiles of the $\mathrm{H{\alpha}}$, $\mathrm{H{\beta}}$ and He I lines. In Appendix~\ref{apCirLines}, we illustrate the emission lines before and after the removal of the photospheric contribution for each night of observation.

The $\mathrm{H{\alpha}}$ and $\mathrm{H{\beta}}$ lines exhibit a redshifted absorption component. This absorption reaches its maximum depth at phase 0.477, indicating that the accretion column crosses our line of sight at this phase. Furthermore, the maximum emission of the He I line at the same phase suggests that the accretion shock region also intersects our line of sight along with the accretion funnel. These results indicate that the accretion column and the accretion shock are spatially correlated with the inner disk warp, which obscures the star around phase 0.5, as observed in the light curves of JH~223.

\subsubsection{Mass accretion rate} \label{sec: Mass accretion rate}

We estimated the mass accretion rate ($\dot{M}_{\mathrm{acc}}$) from the flux of the $\mathrm{H{\alpha}}$ and $\mathrm{H{\beta}}$ lines, which are considered to be good accretion tracers in optical spectra \citep{Gullbring_1998,alcala2017x}. Our spectra were not flux calibrated, so we calculated the line flux ($F_{\mathrm{line}}$) as the product of the line equivalent width ($EW_{\mathrm{line}}$) and the stellar continuum flux at the rotational phase of the spectral observation,
\begin{equation}\label{eq:lineflux}
F_{\mathrm{line}} = EW_{\mathrm{line}} F_{0} \times 10^{-0.4 (m-A)},
\end{equation}
where $F_{0}$, $m$, and $A$ are the reference flux of a zero magnitude star, the magnitude, and the extinction of JH~223 in the corresponding photometric band. We calculated $EW_{\mathrm{line}}$ by integrating the spectrum around the central wavelength of $\mathrm{H{\alpha}}$ and $\mathrm{H{\beta}}$. The results obtained are shown in Table \ref{tab:parhalpha}.

For the $\mathrm{H{\alpha}}$ line, we considered the $R$ band values: $F_{0} = 2.177 \times 10^{-9}$\;erg\;s$^{-1}$\;cm$^{-2}$\;\AA\; \citep{bessell1998model},  $A_R = 0.9 \pm 0.2$ \citep[derived from the extinction relation $A_{R}/A_{V}=0.75$,][]{cardelli1989relationship}, choosing the $R$ band magnitude from LCOGT closest in rotational phase to the spectral observation. For this line we obtained $EW_{\mathrm{H\alpha}}$ ranging from 2.66\;\AA\;to 4.56\;\AA, resulting in line fluxes going from $3.06 \times 10^{-14}$\;erg\;s$^{-1}$\;cm$^{-2}$ to $6.26 \times 10^{-14}$\;erg\;s$^{-1}$\;cm$^{-2}$.

For the $\mathrm{H{\beta}}$ line, we used the $V$ band values: $F_{0} = 3.631 \times 10^{-9}$\;erg\;s$^{-1}$\;cm$^{-2}$\;\AA\;  \citep{bessell1998model}, $A_V = 1.2 \pm 0.2$ \citep{herczeg2014optical}, and chose the $V$ band magnitude from LCOGT closest in rotational phase to the spectral observation. Knowing that the $EW_{\mathrm{H\beta}}$ ranges from 1.42\;\AA\;to 2.64\;\AA, we obtained line fluxes going from $0.91 \times 10^{-14}$\;erg\;s$^{-1}$\;cm$^{-2}$ to $1.81 \times 10^{-14}$\;erg\;s$^{-1}$\;cm$^{-2}$.

We derived the line luminosity from $L_{\mathrm{line}} = 4 \pi d^{2} F_{\mathrm{line}}$, using the stellar distance of $d = 143 \pm 3$\;pc, and calculated the accretion luminosity ($L_{\mathrm{acc}}$) from the empirical relations between $L_{\mathrm{line}}$ and $L_{\mathrm{acc}}$ provided by \citet{alcala2017x}. The accretion luminosity can be converted to mass accretion rate  \citep{Gullbring_1998} via\begin{equation}
    \dot{M}_{\mathrm{acc}} = \left(1-\frac{R_{\star}}{r_{m}}\right)^{-1} \frac{L_{\mathrm{acc}}R_{\star}}{GM_{\star}} ,
    \label{eq:accre_rate}
\end{equation}
where $R_{\star}$ and $r_{m}$ are the stellar radius and the disk truncation radius, respectively. We used $r_{m} = 5$\;R$_{\star}$, which is a typical value used in previous works \citep{Gullbring_1998,bouvier2006magnetospheric,alcala2017x,pouilly2020magnetospheric}. Table \ref{tab:parhalpha} shows the results of the mass accretion rate calculations for $\mathrm{H{\alpha}}$ and $\mathrm{H{\beta}}$. We obtained a mean mass accretion rate of $(5 \pm 1) \times10^{-11}$\;M$_{\odot}$\;yr$^{-1}$ from $\mathrm{H{\alpha}}$ and $(8 \pm 2)\times10^{-11}$\;M$_{\odot}$\;yr$^{-1}$ from $\mathrm{H{\beta}}$. The average mass accretion rate, computed using these values, was $(7 \pm 1)\times10^{-11}$\; M$_{\odot}$\;yr$^{-1}$. \citet{Hartmann16} and \citet{Betti2023} showed that there is a linear trend between mass accretion rate and stellar mass for young stars. When comparing their results and our mass accretion rate values, JH~223 presents a mass accretion rate lower than most young stars of similar mass. 

We also calculated the mass accretion rates using the equivalent widths derived from the color index in the narrowband filters, $CI_{\rm H\alpha}$, as described in Sect.\,\ref{Subsec:OACT_Photometry}. The average mass accretion rate we obtained was $\dot{M}_{\mathrm{acc}} = (5.5 \pm 0.8) \times 10^{-11}$\;M$_{\odot}$\;yr$^{-1}$. These results align with the values calculated from the GRACES and Keck spectra, showing similar variability amplitude and average value. Furthermore, we observed a rotational modulation in the accretion luminosity, with a minimum occurring around phase 0.5 (see Fig.~\ref{fig:lacc_vs_phase}). The modulation observed in $L_{\mathrm{acc}}$ indicates that the line emission region is obscured by the inner disk warp around phase 0.5. When the accretion luminosity is converted to $\dot{M}_{\mathrm{acc}}$, it would imply that the accretion rate is lower exactly when the main accretion stream, which creates the inner disk warp (see Sect. \ref{sect:discussion}), is expected to be facing the observer. Therefore, we emphasize that a decrease in $L_{\mathrm{acc}}$ does not necessarily correspond to a real decrease in $\dot{M}_{\mathrm{acc}}$, but it instead reflects the obscuration of the accretion flow at this phase.

\subsection{Near-infrared spectropolarimetry}
\subsubsection{Veiling}\label{veiling}

The nIR veiling may have a distinct origin compared to the optical veiling. It is caused by excess emission from the circumstellar disk, resulting from dust heating in its inner regions. This heating is associated with reprocessing stellar radiation and releasing energy during the accretion process \citep{Chiang1997,2023arXiv230102450S}.

To measure the nIR veiling of JH~223, we applied the same procedures as described in Sect. \ref{sec:optical_veiling}, using the SPIRou spectra of JH~223. We selected the spectral region between $1158$\;nm \ and $1167$\;nm, in the $Y$ band, to measure the veiling as it is close to the He I 1083\;nm line, which is one of the lines of interest in our analysis. As a template, we used the SPIRou spectra of the weak-line T Tauri star TWA 7.

\begin{table}[ht]
\caption{Log of the SPIRou observations of JH~223. }
\centering
\begin{tabular}{ccccc}
\hline
\hline \
 $E$ &  Veiling & $RV$ & $B_\mathrm{l}$ & $\sigma_\mathrm{LSD}$ \\
  &    & (km/s) & (G) &  ($10^{-4}$)    \\
\hline
 6.278  & $0.09 \pm 0.02$ &  $16.2 \pm 0.2$   & $-139 \pm 16$  & 4.38   \\
 7.499  & $0.15 \pm 0.02$ &  $15.5 \pm 0.3$ & $-147 \pm 19$  & 5.43      \\
 7.808  & $0.13 \pm 0.03$ &  $15.2 \pm 0.3$ &  $0 \pm 15$ &  6.68   \\
 8.068  & $0.20 \pm 0.03$  & $15.4 \pm 0.3$  &  $-136 \pm 19$  & 5.52  \\
 8.422  & $0.00 \pm 0.02$ &  $15.3 \pm 0.3$ &  $-190 \pm 24$ & 6.12    \\
 8.701  & $0.00 \pm 0.02$ & $14.2 \pm 0.3$  & $-87 \pm 23$  & 9.73    \\
 9.284  & $0.10 \pm 0.02$ & $15.9 \pm 0.2$  & $-124 \pm 16$  &  4.82  \\
 9.591  & $0.15 \pm 0.02$ &  $14.8 \pm 0.3$ & $-60 \pm 13$  & 4.56    \\
\hline
\end{tabular}
  \label{tab:results_spirou}
  \tablefoot{Column 1 lists the rotational cycle ($E$), Col.~2, the average veiling calculated in the region from $1158$\;nm to $1167$\;nm, Col.~3 shows the stellar radial velocity, Col.~4 presents the longitudinal magnetic field, and Col.~5, the median of the error of Stokes $V$ LSD profiles.}
\end{table}

Table \ref{tab:results_spirou} shows the veiling computed for each night of observation. We derived the average veiling of $0.10 \pm 0.07$, where the error was calculated using the standard deviation of all measurements. We did not detect any rotational modulation in the veiling variability. We conclude that our low veiling values are in agreement with the optical veiling results obtained in Sect. \ref{sec:optical_veiling}. We also measured the veiling in the spectral region from $2260$\;nm to $2269$\;nm, in the $K$ band. The average veiling value of $0.28 \pm 0.07$ was larger than in the $Y$ band, a behavior consistent with the results reported by \citet{2023arXiv230102450S}, and the veiling values showed, once more, no rotational modulation.

\subsubsection{Emission lines}\label{sec:Infrared_emission_lines}

We analyzed the He I 1083\;nm and $\mathrm{Pa}\beta$ lines present in the SPIRou spectra of JH~223. The He I line may originate in the accretion funnel, the accretion shock, and winds \citep{erkal2022he}, while $\mathrm{Pa}\beta$ is thought to be formed mostly along the accretion funnel. Therefore, these lines are good tracers in the infrared of accretion and mass loss \citep{2021A&A...649A..68S}. We did not detect any other circumstellar line exhibiting signs of accretion.

We removed the photospheric contributions from the JH~223 spectra in the  He I and $\mathrm{Pa}\beta$ line regions to obtain only the circumstellar contribution. Thus, we subtracted  a veiled and rotationally broadened mean spectrum of TWA 7 from the JH~223 spectra. 

\begin{figure}
    \centering
     \includegraphics[width=0.8\linewidth]{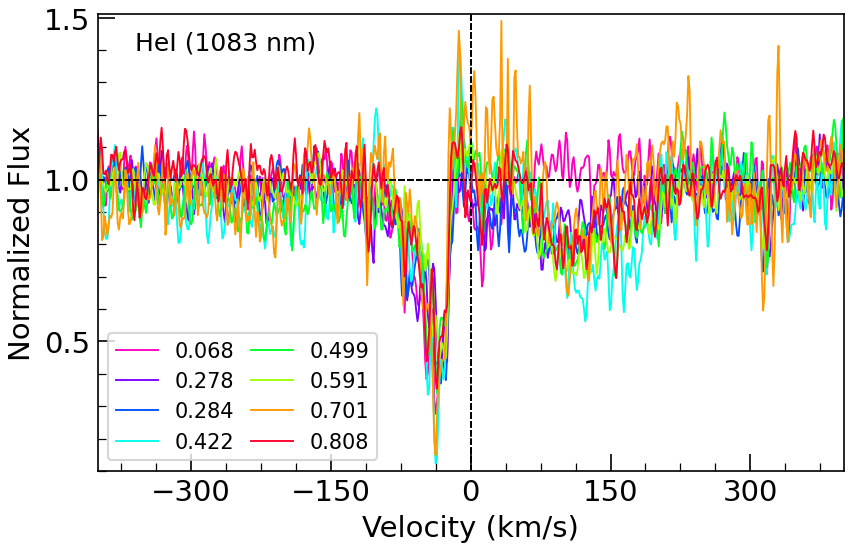}
    \caption{Residual line profiles of He\;I~1083\;nm. The colors represent different rotational phases.}
    \label{fig:HeI-overplot}
\end{figure}

The circumstellar $\mathrm{Pa}\beta$ line did not show any relevant signal, indicating that the absorption lines in the observed spectra were all photospheric (see Fig. \ref{fig:residuals-SPIRou}). The circumstellar He I line presents blueshifted and redshifted absorption components, as shown in Fig. \ref{fig:HeI-overplot}. The blueshifted absorption component, around $-40$\;km\;s$^{-1}$, is commonly attributed to outflows projected on our line of sight, such as stellar and disk winds, which absorb photons emitted by the star and the accretion funnel \citep{Edwards06,erkal2022he}. This component persists in all observations and shows small variability in depth and width. The redshifted absorption component traces gas in the accretion funnel, moving away from the observer, which absorbs photons from the shock region \citep{fischer2008redshifted}. In the JH 223 spectra, this component exhibits large variations in velocity and depth. It becomes stronger at phases between 0.4--0.6, which correspond to the photometric minima and to the passage of the accretion funnel in our line of sight, than around phases 0.0--0.3, where it is hardly seen (see Fig. \ref{fig:residuals-SPIRou}). This indicates a spatial association between the funnel flow and the inner disk warp.

\subsubsection{Least-squares deconvolution}\label{LSD}

We analyzed the SPIRou spectra by applying the least-squares deconvolution (LSD) method \citep{1997MNRAS.291..658D} to atomic photospheric lines to derive weighted average Stokes $I$ and $V$ profiles for each night of observation. This method increases the $S/N$ of the Stokes parameters \citep{1997MNRAS.291..658D}. In particular, it is well suited for analyzing the Stokes $V$ profiles, which usually present very low S/N values in the individual photospheric lines of T Tauri stars \citep{1997MNRAS.291..658D, kochukhov2010least}. 

Overall, LSD assumes that the intensity (polarization) spectrum results from a convolution between the weighted mean Stokes $I$ (Stokes $V$) LSD profile and a mask of photospheric atomic lines.  We used the VALD database and MARCS atmosphere models to generate the line mask, considering an effective temperature of $T_\mathrm{eff} = 3500$\;K, surface gravity of $\log {g} = 4.0$, and microturbulence velocity of $v_\mathrm{mic} = 2.0$\;km/s. These values are compatible with the stellar parameters of JH~223 (see Sect.~\ref{stellar_param}). We further selected only atomic lines with known Landé factors and a depth larger than $3\%$ of the continuum level, resulting in a final mask with 2307 photospheric lines. 

The LSD profiles were computed using the \verb|LSDpy| code with a window of $200$\;km/s, adopting for the normalization the average Landé factor of 1.2, a central wavelength of 1700\;nm, and a line depth of 0.15. In the next sections, we use the photospheric Stokes $I$ and $V$ LSD profiles, shown in Fig. \ref{fig:stokesIV}, to investigate the radial velocity and longitudinal magnetic field variability, and perform a Zeeman-Doppler imaging analysis of the spectropolarimetric data of JH 223.

\subsubsection{Radial velocity}\label{sec:RV}
\begin{figure}
    \centering
    \includegraphics[width=0.9\linewidth]{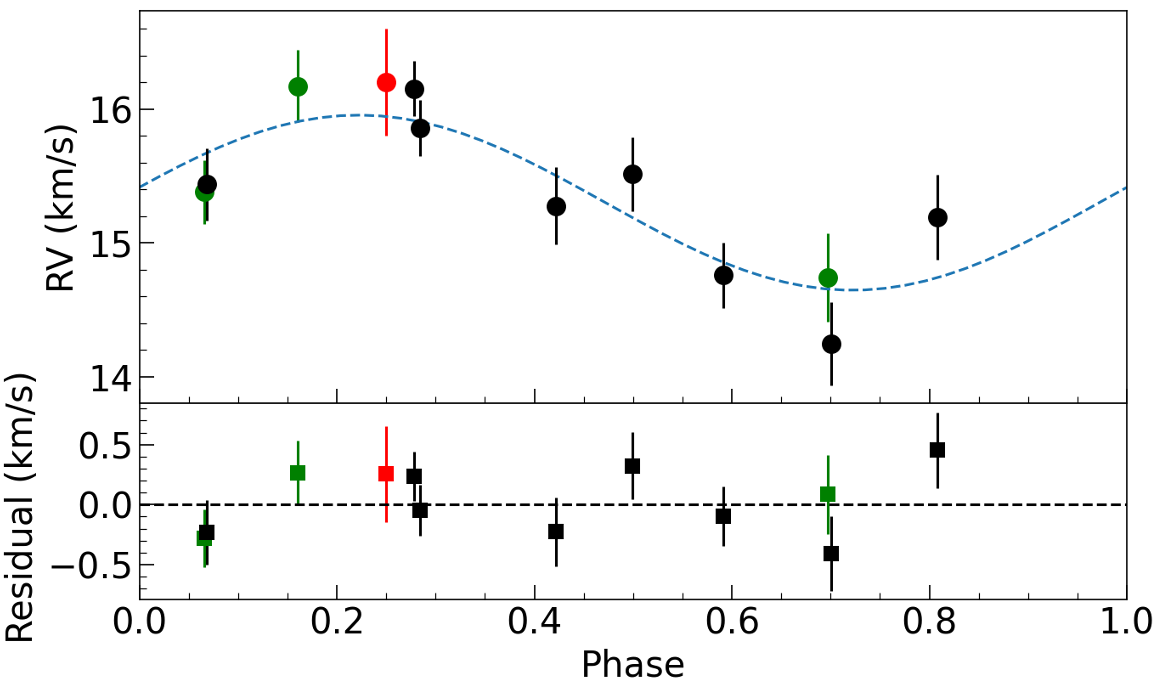}
    \caption{JH~223 radial velocities (top) and residuals (bottom). Radial velocities were derived from the first-order moment of the Stokes $I$ LSD profiles of the SPIRou observations (black dots) and from spectral fitting of GRACES (red dot) and Keck spectra (green dots). The blue curve represents the sinusoidal fit with the rotational period fixed at $3.31$\;days. Residuals (squares), computed as the difference between the RV values and the sinusoidal fit, have a standard deviation of $0.27$\;km/s.}

    \label{fig:vr_stokes}
\end{figure}

Radial velocity values were calculated as the first-order moment of the Stokes $I$ LSD profiles, as in \citet{2017MNRAS.465.3343D}, in the interval of $\pm 30$\;km/s with respect to the profile center. The radial velocities calculated for each night of observation are listed in Table \ref{tab:results_spirou} and shown in Fig.~\ref{fig:vr_stokes}, as a function of the rotational phase using Eq.~\ref{ephemeris}. We obtained radial velocities ranging from $14.2$\;km/s to $16.2$\;km/s, with an average value of $15.3 \pm 0.6$\;km/s. The radial velocity variation corresponds to distortions of the Stokes $I$ LSD profiles due to the presence of spots at the stellar surface, while the mean radial velocity value is attributed to the true radial velocity of JH~223. We calculated the error of individual radial velocities propagating the uncertainties in the first-order moment equation of Stokes $I$. We obtained the error of the average value from the standard deviation of the individual measurements. The radial velocities obtained with the GRACES and Keck spectra (see Sect.~\ref{stellar_param}) agree within $1 \sigma$ with the individual radial velocity measurements from SPIRou (see Fig.~\ref{fig:vr_stokes}).

To verify whether the modulation of the radial velocity caused by spots on the stellar surface is associated with the period derived from the K2 light curve, we phase-folded the RV measurements using the ephemeris in Eq.~\ref{ephemeris}. We then fit a sinusoidal curve to the phase-folded data, fixing the period to that of the light curve and optimizing the amplitude and phase offset parameters of the sine function. The fit showed good agreement between the RV modulation and the K2 light curve period, suggesting the stellar rotation period is consistent with the rotational period of the warp in the inner disk. This result indicates that the disk deformation responsible for the dips in the K2 light curve is located close to the disk's co-rotation radius. Furthermore, the negligible phase offset of the sine function (0.02 $\pm$ 0.05) suggests that the main spot on the stellar surface crosses our line of sight around phase 0.5 (see Sect. \ref{ZDI} and Fig. \ref{fig:bright_polar}), which is in agreement with the warp location. It is important to bear in mind that we cannot discard the possibility that the warp itself induces part of the RV variations, as it modulates the visible area of the stellar surface, much as a cool spot would do.

\subsubsection{Longitudinal magnetic field}\label{sec:bl}

\begin{figure}
    \centering
    \includegraphics[width=0.9\linewidth]{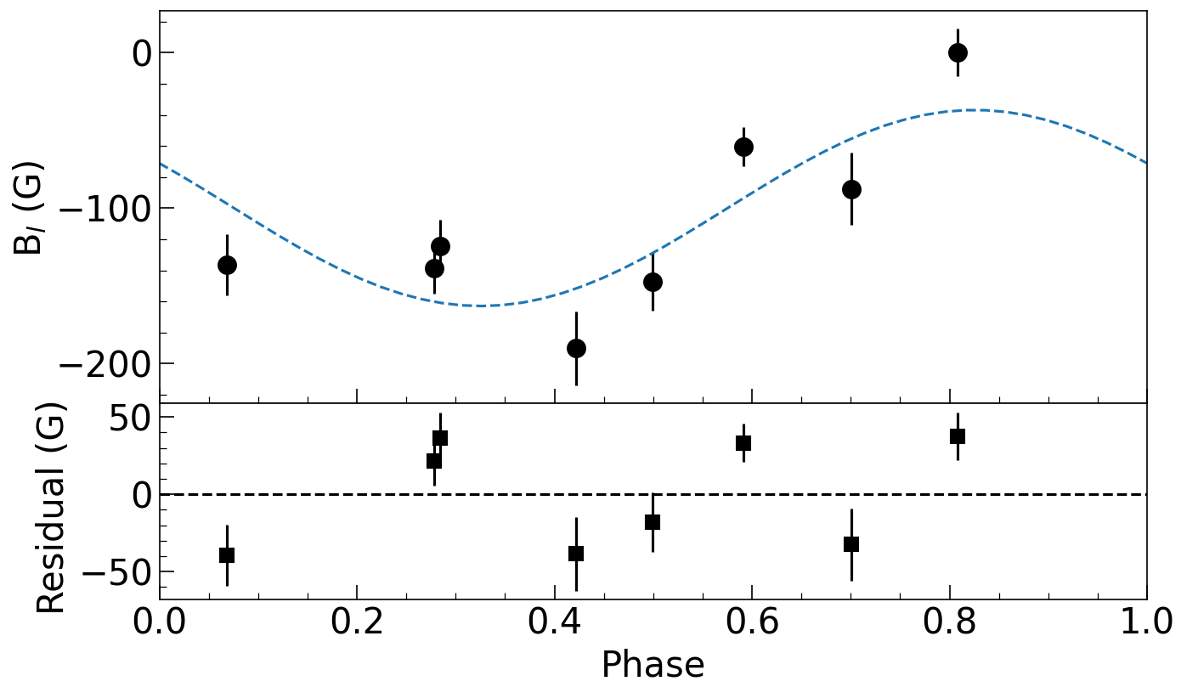}
    \caption{Same as Fig.~\ref{fig:vr_stokes}, but for the longitudinal magnetic field. Residuals have a standard deviation of $33$ G.}
    \label{fig:bl_stokes}
\end{figure}

We estimated the longitudinal magnetic field component ($B_\mathrm{l}$) from the first-order moment of the Stokes $V$ LSD profiles  \citep{kochukhov2021magnetic} via
 \begin{equation}
 B_{\ell} =  -\frac{7.145 \times 10^{5}}{\lambda_{0}g_\mathrm{eff}}  \frac{\int [v-\upsilon_\mathrm{rad}]V(v)dv}{\int [1-I(v)]dv},
 \label{blong}
 \end{equation}
where $B_{\ell}$ is expressed in G, $\lambda_{0}$ is the average wavelength in nm, and $v$ and $\upsilon_\mathrm{rad}$ denote the velocity at each point along the Stokes $V$ profile and the radial velocity of the star, respectively, both in km/s. The average wavelength, $\lambda_{0}$, and the effective Landé factor, $g_\mathrm{eff}$, are the same ones used in Sect.~\ref{LSD} to compute the LSD profile. $I(v)$ and $V(v)$ correspond to the Stokes $I$ and $V$ LSD profiles, respectively. We computed equation \ref{blong} in the $\pm 30$\;km/s interval with respect to the center of the Stokes $I$ LSD profile. We obtained a longitudinal magnetic field ranging from $0$\;G to $-190$\;G with an average field value of $-111 \pm 56$\;G. The values obtained for each observation are presented in Table \ref{tab:results_spirou} and shown as a function of the rotational phase in Fig.~\ref{fig:bl_stokes}. We fitted a sine function to the $B_{l}$ values, keeping the period of $3.31$\;days fixed, and found a modulation in agreement with the rotational period of JH~223.

\subsubsection{Zeeman-Doppler imaging}\label{ZDI}

We used the tomographic technique of Zeeman-Doppler imaging \citep[ZDI, ][]{semel1989zeeman} to obtain information on brightness and magnetic field features at the surface of JH~223. This technique consists of tracing rotationally modulated signatures present in the time-series Stokes $I$ and $V$ LSD profiles to reconstruct the brightness distribution and large-scale magnetic field topology at the stellar surface \citep{1997MNRAS.291..658D,donati2006surprising}. For this purpose, we employ the \verb|ZDIpy| code written by \citet{folsom2018evolution}, which has been largely used in the literature to model active stars \citep{cang2020magnetic, pouilly2021beyond,bellotti2023monitoring}.

The ZDI model treats the star as a spherical surface divided into a grid with elements of approximately equal area. We configured the surface of our model with $60$ longitudinal rings, which correspond to approximately $2000$ surface elements. The poles of the spherical coordinate system are aligned with the star's rotation axis. The algorithm computes synthetic local Stokes $I$ profiles for each surface element by employing a Voigt profile, weighted by its projected area, limb darkening, and local brightness. We derived the Voigt profile parameters (specifically, the Gaussian and Lorentzian widths) through a fitting procedure to the Stokes $I$ LSD profile of the weak-line T Tauri star TWA7 (see Sect. \ref{sec:optical_veiling}), a slow rotator of the same spectral type as JH~223.  The resultant fit provided Gaussian and Lorentzian widths of $9.7$ km/s and $0.06$ km/s, respectively. We retrieved the linear coefficient of limb darkening from the tables of \citet{claret2011gravity}, using the value of $0.2865$ for TWA 7 and $0.2792$ for JH~223. The local Stokes $V$ profiles are derived using a weak field approximation, expressed as a function of the derivative of the Stokes $I$ profile \citep{folsom2016evolution}, and the magnetic field is represented as a combination of spherical harmonics for the radial, azimuthal, and meridional components. The full disk-integrated line profiles, computed as the summation of local profiles at the visible surface Doppler-shifted by the line-of-sight projected velocity of the surface element, are used to reconstruct the surface brightness or the magnetic field at the stellar surface. The brightness map starts with a homogeneous brightness distribution, and the code iteratively adds dark and bright spots to the stellar surface to identify the optimal brightness distribution that reproduces the observed Stokes $I$ profiles. To optimize the modeling of the large-scale magnetic field, the code adjusts the spherical harmonic coefficients of the radial, meridional, and azimuthal field components. This iterative process applies the principles of maximum entropy \citep{skilling1984maximum} and $\chi^{2}$ minimization as regularization criteria to the image reconstruction. 

In our preliminary analysis, the brightness and magnetic field maps converged rapidly to  $\chi^{2}_{r} \sim 1$, resulting in models with little information. \citet{wade2000high,wade2000spectropolarimetric} argued that the error bars of the LSD profiles are typically overestimated, bringing the need to aim for a target $\chi^{2}_{r}$ lower than 1 to compute the amount of information suitable for brightness and magnetic field maps. This highlights the need for a robust method to select the target $\chi^{2}_{r}$ that allows the ZDI model to extract information without overfitting the data. We proposed to apply a scaling factor to the error bars of the Stokes $I$ and $V$ profiles, in a similar way to that done by \citet{wade2000spectropolarimetric}, following
\begin{equation}\label{eq:newsigma}
 \sigma_{new}=  \sigma_{LSD}\sqrt{\chi^{2}_{r,sf}},
\end{equation}
where $\sigma_{LSD}$ is the LSD error and  $\chi^{2}_{r,sf}$ is the value that extracts information from the data without overfitting. As detailed in Appendix~\ref{apChi}, we obtained $\chi^{2}_{r,sf} = 0.53$ for our dataset. We applied the correction to the error bars of all LSD profiles, which allowed us to target a unit $\chi^{2}_{r}$ in the following image reconstructions.

\begin{figure}
    \centering
\includegraphics[width=0.55\linewidth]{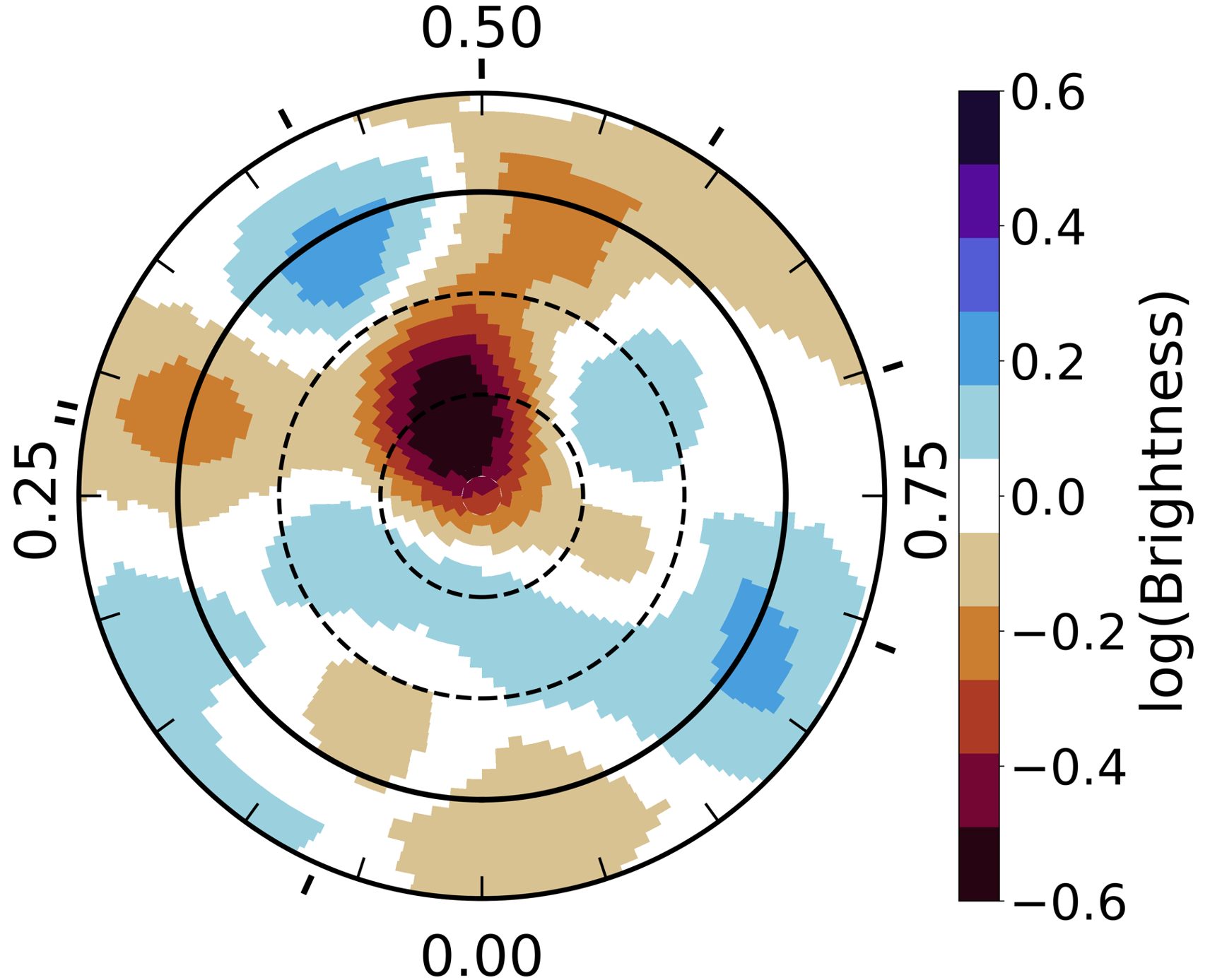}
    \caption{Logarithmic brightness map of the surface of JH~223 in November, 2019. The star is shown in a flattened polar view down to a latitude of $-30^{^\circ}$, with the equator indicated by a solid line and latitudes of $60^{\circ}$ and $30^{\circ}$ by dashed lines. Outer radial ticks mark the phases of spectropolarimetric observations. Cool spots are shown in brown shades, while bright plages are shown in blue shades.}
    \label{fig:bright_polar}
\end{figure}

After applying the procedure described in the previous sections, we modeled the brightness and large-scale magnetic field topology on the surface of JH~223. Initially, we reconstructed the brightness distribution fitting the observed Stokes $I$ LSD profiles (Fig. \ref{fig:stokesIV}, left). The reconstructed brightness map is displayed in Fig.\;\ref{fig:bright_polar} with a spot and plage coverage of $18\%$ of the stellar surface. The map shows a high contrast cool spot near the pole, located at about 60$^\circ$ latitude with respect to the equator, toward phases $0.40-0.45$. Additionally, there are lower contrast features near the equator.

\begin{figure}
    \centering
    \includegraphics[width=0.54\linewidth]{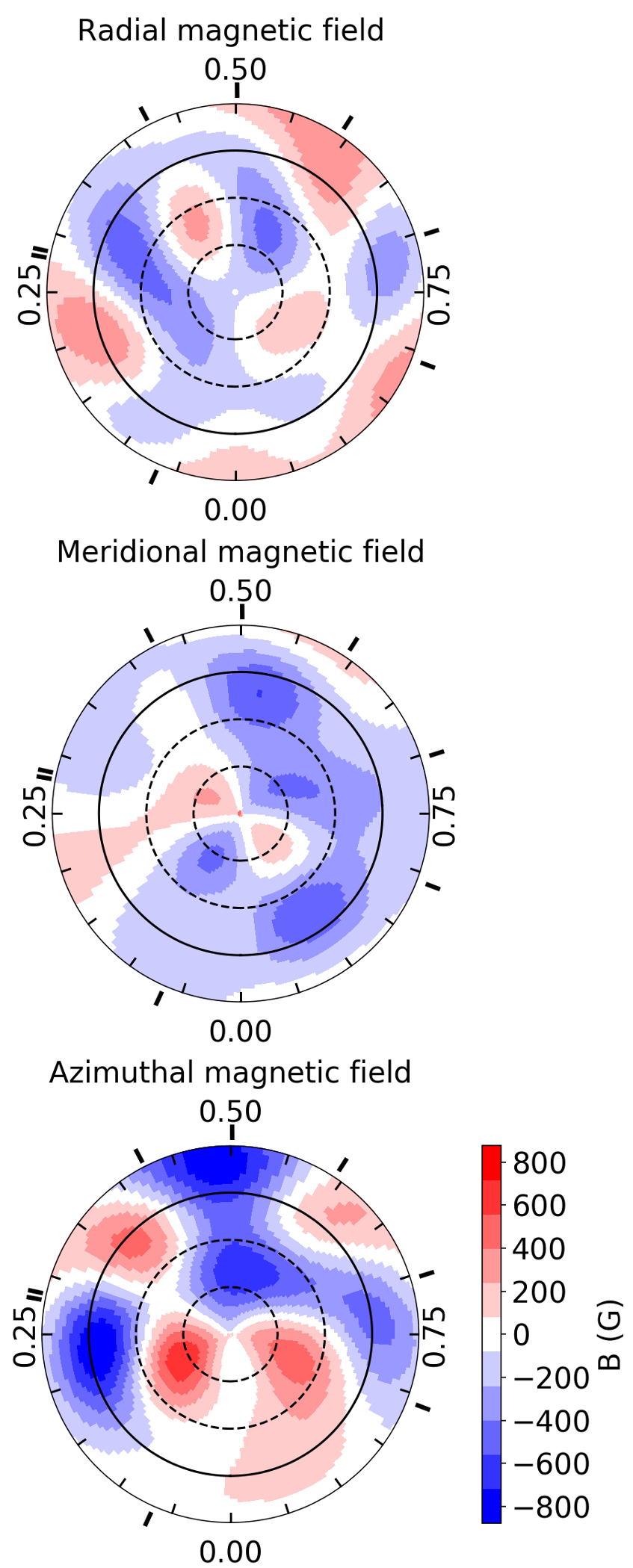}
    \caption{ZDI maps of the radial (top), meridional (middle), and azimuthal (bottom) components of the large-scale magnetic field at the surface of JH~223. Similar to Fig.\ref{fig:bright_polar}, the star is represented in a flattened polar projection. Magnetic fluxes, indicated by the color bar, are expressed in gauss.}
    \label{fig:magfieldpolar}
\end{figure}

We then modeled the time series Stokes $V$ profiles (Fig. \ref{fig:stokesIV}, right) to obtain the large-scale magnetic field topology, imposing the brightness distribution previously calculated. We truncated the spherical harmonics expansion describing the magnetic field at the order of $\ell_{max} = 5$. Including terms with a higher order did not significantly change the reconstructed magnetic field, in line with what is expected for stars with low $v\sin {i}$ \citep{morin2010large}. The resulting maps for radial, azimuthal, and meridional components of the magnetic field are shown in Fig.~\ref{fig:magfieldpolar}. The large-scale magnetic field topology is predominantly poloidal, with $71\%$ of the total magnetic energy, while the toroidal component represents $29\%$ of the total magnetic energy. The poloidal field has $28\%$ of axisymmetry, and the dipolar component is the most significant component, with $37\%$ of the poloidal energy, tilted by $28^{\circ}$ toward phase 0.38. The quadrupolar and octupolar components represent $6\%$ and $14\%$ of the poloidal energy, respectively. The average unsigned magnetic field strength across the stellar surface is $358$\;G, with the maximum field intensity reaching $915$\;G. The magnetic properties of JH 223 are summarized in Table \ref{tab:magnetic-field-properties}.

\begin{table}[]
\caption{Magnetic field properties of JH~223 derived with ZDI.}
\centering
\begin{tabular}{ll}
\hline
\hline
Magnetic field properties   &  \\ \hline
$B_\mathrm{mean}$ (G)                 & 358 \\
$B_\mathrm{max}$ (G)                 & 915 \\
$E_\mathrm{pol}$ ($E_\mathrm{Tot}$)           & 71\% \\ 
$E_\mathrm{Tor}$ ($E_\mathrm{Tot}$)           & 29\% \\ 
$E_\mathrm{Dip}$ ($E_\mathrm{Pol}$)           & 37\% \\ 
$E_\mathrm{Quad}$ ($E_\mathrm{Pol}$)          & 6\% \\ 
$E_\mathrm{Oct}$ ($E_\mathrm{Pol}$)           & 14\% \\ 
Pol. axisymmetric ($E_\mathrm{pol}$)          & 28\% \\ 
$B_\mathrm{Dip}$ (G)                 & 250 \\ 
$\beta (^{\circ})$                   & 28 \\ 
Rot. phase tilt             & 0.38 \\ \hline
\end{tabular} \label{tab:magnetic-field-properties}
\end{table}

\section{Discussion}\label{sect:discussion}

JH~223 is a classical T Tauri star that exhibited typical dipper-like behavior when it was observed by K2 in 2017. The K2 light curve shows a constant maximum flux interrupted by minima that vary in width and depth from cycle to cycle. The $Q$ and $M$ parameters, calculated following the methodology of \citet{Cody14} and \citet{Cody18}, further support the classification of this light curve as a typical dipper, indicating a clear quasiperiodic variability ($Q = 0.37$) with positive asymmetry ($M = 0.76$). This quasiperiodic variability is similar to the photometric variability of known dipper stars, such as AA Tau \citep{Bouvier07a} and LkCa 15 \citep{2018A&A...620A.195A}, and indicates the presence of circumstellar dust in the inner disk region that occults the star as the system rotates. These dust occultations are thought to come from an inner disk warp produced by the interaction of an inclined stellar dipole magnetic field component and the inner circumstellar disk, as seen in MHD simulations \citep{2013MNRAS.430..699R}. Our observing photometric, polarimetric, and spectroscopic campaign was carried out at the end of 2019, with the ground-based light curves obtained displaying the same characteristics as those seen in the prior K2 data. The CMDs show that the system becomes redder when fainter, and the optical linear polarimetric observations present an increase in the degree of polarization in the photometric minima, which gives support for a dust obscuration origin of the photometric variations. 

JH~223 was later observed with the TESS satellite in 2021 and 2023. The 2021 TESS light curve (sectors 43 and 44; see Fig. \ref{fig:tess4344}) shows a much more regular variability pattern, reminiscent of spot-like light curves \citep[see][]{Cody18}, but with a scalloped morphology, as though some dust still obscured the star at that epoch. Although the flux asymmetry for these sectors still corresponds to a dipper-like light curve with $M = 0.5$, the periodicity parameter $Q = 0.04$ suggests that a periodic physical process, such as spots, dominates the light curve. The influence of spots on the light curve variability is significant in 2021, imposing a strict periodicity that is typically not observed in dipper-like stars. This finding indicates that the presence of cool spots on the stellar surface, modeled with ZDI in 2019, is recurrent in JH 223. Additionally, there is a decrease of about 20$\%$ in the amplitude of the flux variability in the 2021 TESS light curve when compared to the other epochs, suggesting a lower amount of circumstellar dust at the warp location.

\begin{figure}
    \centering
    \includegraphics[width=0.9\linewidth]{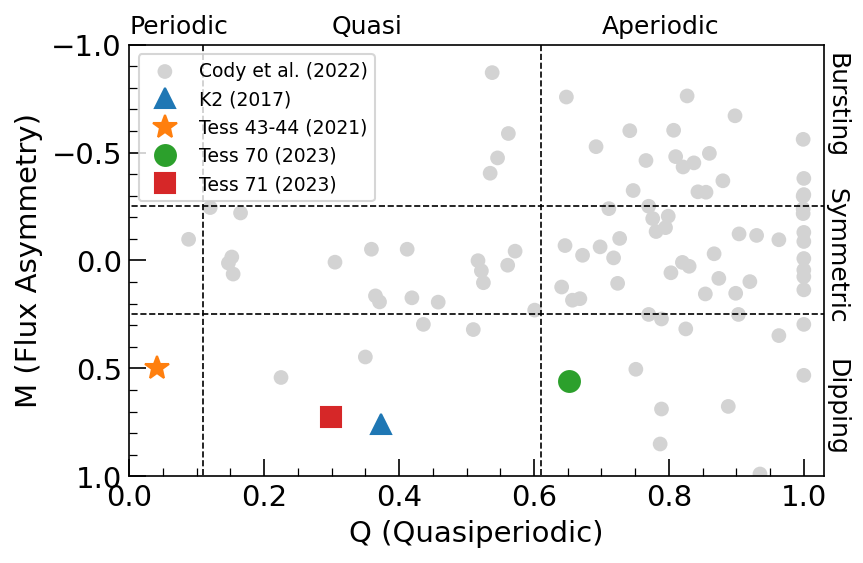}
    \caption{Asymmetry ($M$) and quasiperiodicity ($Q$) parameters derived from the K2 and TESS light curves of JH~223. The gray points correspond to the $Q$ and $M$ parameters of others CTTSs analyzed by \citet{Cody22} in the Taurus star-forming region.}
    \label{fig:Q-Mcody}
\end{figure}

The 2023 TESS light curve variability seemed again to be due mostly to obscuration by dust (sectors 70 and 71, see Fig. \ref{fig:tess7071}). However, sector 70 presented an asymmetric ($M = 0.56$) and aperiodic ($Q = 0.65$) light curve, as though the circumstellar dust were randomly distributed in the inner disk region, as predicted by unstable+ordered and solely unstable MHD models \citep{2016MNRAS.459.2354B}. Remarkably, the phased light curve of Sector 70 suggests two main obscuration events occurring at intermediate phases of $\sim\!0.3$ and $\sim\!0.8$; namely, in quadrature, compared to the main dip seen at phase 0.5 at other epochs. After about 25 days of observations, the photometric light curve in sector 71 gradually recovered the quasiperiodicity ($Q = 0.30$) and asymmetry ($M = 0.73$) that are typical of dipper-like stars, and looked again very similar to the K2 and our ground-based data. The morphological light curve variation in several timescales over the years seems to indicate that the dust distribution in the inner disk changes considerably in timescales of weeks to years. \citet{2016MNRAS.459.2354B} MHD simulations showed the development of a pair of opposite accretion tongues in the ordered unstable regime, which, at times, seems to be perpendicular to the longitude of the magnetic pole (see their Fig.9). We might thus have witnessed the system transitioning from an ordered unstable regime during TESS Sector 70 (with two opposite tongues nearly perpendicular to the magnetic pole, producing the dips at phases 0.3 and 0.8) and going back to a dominant funnel flow in Sector 71, along with the associated disk warp responsible for the main dip at phase 0.5. Figure \ref{fig:Q-Mcody} shows the variability of the $Q$ and $M$ parameters for the light curves of JH 223 over six years of observations, overplotted with the parameters obtained for others CTTSs with confirmed disks in the Taurus star-forming region \citep{Cody22}. Similar variations in light curve morphology of CTTSs on a timescale of 3 years were already pointed out by \citet{Mcginnis15} when analyzing CoRoT satellite light curves of 84 CTTSs in NGC 2264. They showed that about half of the dipper-like light curves became aperiodic and vice-versa on this timescale, hinting at the highly dynamical nature of the star disk interaction that shapes the dust distribution in the inner disk of accreting systems. This type of photometric behavior may actually be typical of low-mass accretion rate systems, as it was also reported for  PDS 70 \citep{Gaidos24}, a K7 CTTS with a transition disk and two orbiting exoplanets, which was observed with TESS on three different epochs over four years and showed light curves that varied from spot to dust dominated. PDS 70 also presented $\mathrm{H{\alpha}}$ and He I 1083 nm line profiles similar to those observed in JH~223 at certain epochs \citep{Thanathibodee20, Gaidos24}, showing that accretion and outflow are ongoing with about the same characteristics in both systems. As JH~223 is younger than PDS 70 and lacks detected planets, we may be witnessing the early stages of planet formation, potentially triggered by disk instabilities that could explain the observed dust density variability.

Despite the morphological changes in the light curve, the periodicity of the photometric variability of the JH 223 TESS observations was maintained most of the time in the almost seven years of photometric data analyzed in this work. It is important to keep in mind that the period measured from a dipper light curve actually corresponds to the period of the disk radius where the dust overdensity is located, and may not be related to the stellar rotation period. However, in the case of JH~223 (as with many other dipper CTTSs), the inner disk warp at $r_\mathrm{m}=6 \pm 1$\;R$_\mathrm{\star}$, calculated with Equation (6) of \citet{Bessolaz08}, occurs at the corotation radius between the star and the disk, $r_\mathrm{cor}=6 \pm 1$\;R$_\mathrm{\star}$. The stellar rotation period (and, consequently, the location of the dust at the corotation radius) can also be verified through the radial velocity and longitudinal magnetic field variations, which come from cold spots at the stellar photosphere and the surface magnetic field, which also vary at the same period as the K2 light curve. The fact that the 2021 TESS light curve seems to be dominated by spots rather than dust variability, still showing the same periodicity as before, also points to the location of the circumstellar dust warp, when present, near the corotation radius. Additionally, the dust sublimation radius, calculated from Eq. 1 of \citet{Monnier02} with $T_\mathrm{sub}$ = 1500\;K and $Q_{R}=1$, lies at a distance of $r_\mathrm{sub}=2.8 \pm 0.3$\;R$_\mathrm{\star}$, reinforcing that, at the disk warp location ($6 \pm 1$\;R$_\mathrm{\star}$), the disk is sufficiently cold for dust to exist. 

JH~223 presents a low mass accretion rate, with consistent measurements in the optical and in the nIR. The $\mathrm{H{\alpha}}$ and $\mathrm{H{\beta}}$ emission lines, as well as the He I 1083 nm line, show redshifted absorption around phase 0.5, indicating that the main accretion funnel is in our line of sight at about this phase. Furthermore, the maximum emission of the He I 587.6 nm line at the same phase suggests that the accretion shock region also intersects our line of sight along with the accretion funnel. These results coincide with the observations of the quasiperiodic luminosity dips in the contemporaneous LCOGT light curve and show the spatial connection between the accretion shock, the accretion column, and the dust warp in the inner disk.

The reconstructed large-scale magnetic field is complex and presents an inclined dipole component, which creates a major accretion column in each hemisphere, according to MHD simulations of star-disk interaction \citep{2013MNRAS.431.2673K}. The dipole, with a maximum inclination around phase 0.4, is thought to give rise to the warp of gas and dust at the base of the accretion funnel near the disk, which periodically occults the stellar photosphere, with a maximum obscuration near phases 0.4--0.5, as observed. The $\sim$\;0.1 phase lag between the longitude of the magnetic pole and the maximum obscuration from the warp is predicted by analytical models of misaligned magnetic star-disk interaction to result from the delayed response of the inner disk to the magnetic perturbation \citep{Terquem&Papaloizou}. These models also predict a trailing warp, which may account for the asymmetric shape of the main dip at the time of our observations, with a sharper ingress and a shallower egress.

\begin{figure*}
    \centering
    \includegraphics[width=0.9\linewidth]{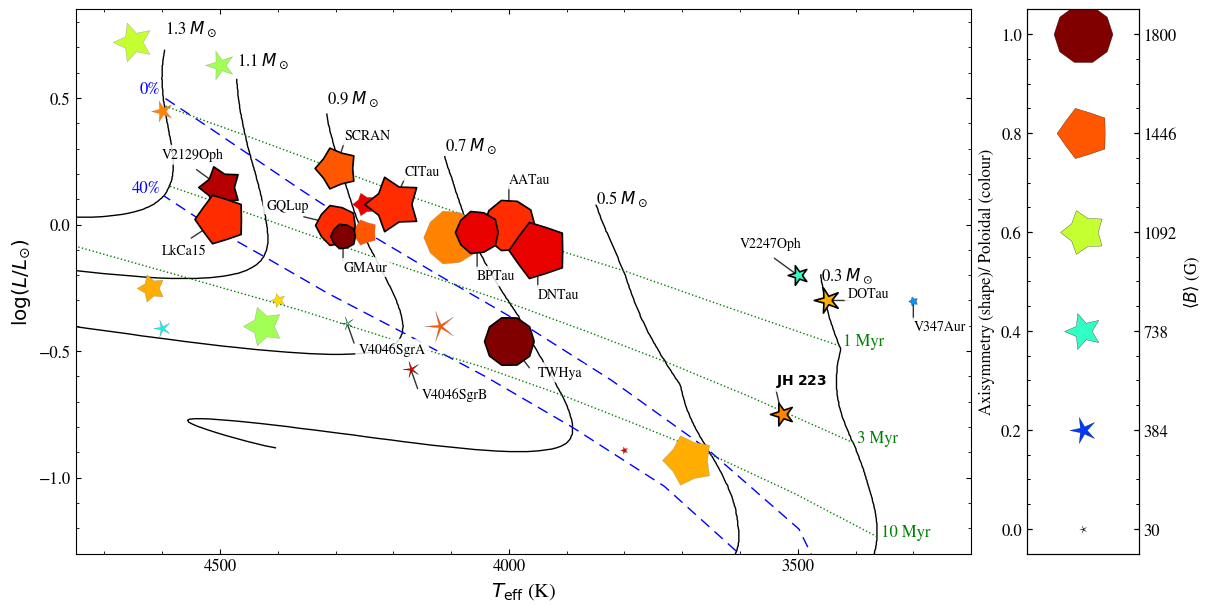}
    \caption{HRDs showing the large-scale magnetic properties of JH~223 and other classical and weak-line T Tauri stars.  CTTSs are labeled by name. Symbol size scales with the average magnetic field strength, color with the poloidal magnetic energy fraction, and shape with the fraction of poloidal energy in axisymmetric modes. Evolutionary tracks (black lines) and isochrones (green dotted lines) from \citet{baraffe2015new} are shown, together with the fully convective boundary and the $40\%$ radiative-core mass limit (dashed blue lines). Magnetic field data for the T Tauri stars are taken from \citet{donati2014modelling,donati2015magnetic,yu2017hot,hill2017magnetic,nicholson2018surface,hill2019magnetic,yu2019magnetic,nicholson2021surface,xiang2023magnetic, donati2007magnetic,donati2008magnetospheric,donati2010magnetospheric,donati2010complex,donati2011large,donati2011close,donati2012magnetometry,donati2013magnetospheric,donati2019magnetic,donati2020magnetic,Zaire24,Donati26}.}
    \label{fig:confusogram}
\end{figure*}

According to our analysis, the characteristic dimming patterns observed in the light curves of JH~223 can be explained by the obscuration of the stellar surface by an inner disk warp, which originates from the interaction of an inclined dipole and the inner disk region. The disk warp is spatially connected to the accretion funnel, as evidenced by the appearance of redshifted absorption components in the \(\mathrm{H{\alpha}}\) and He I 1083 nm lines, generated along the accretion funnel, which coincide in phase with the warp obscuration of the stellar surface. Other scenarios have been proposed in the literature to explain the dipper light curves, such as vertical hydrodynamic disk instabilities or dusty disk winds \citep{Ansdell20}. Hydrodynamic instabilities can form vortices in the inner part of the disk, lifting dust above the disk midplane. These vortices are generated in regions of low ionization and weak coupling between the magnetic field and the gas (dead zones) \citep{Lyra12,Flock17}, and for stars with an inclination between 70$^{\circ}$ and 80$^{\circ}$, this mechanism can be a potential source of overdensity formation that periodically obscures the stellar photosphere \citep{Ansdell20}. In the case of JH~223, the dust overdensity that obscures the star is located at the corotation radius of the disk, which coincides with the disk truncation radius, indicating that the deformation periodically observed is physically associated with the accretion column and, consequently, with the magnetic field, and therefore is not located in a dead zone region. The dusty disk wind scenario can also be excluded since the disk wind in JH~223, detected through a narrow blue-shifted absorption component in the He I line 1083 nm, does not vary with the stellar rotational period, remaining almost constant in our observations, which indicates that the winds mostly originate from a region far from the corotation radius.

Figure~\ref{fig:confusogram} shows JH~223 in an HRD that displays the large-scale magnetic characteristics of young low-mass stars, reconstructed with ZDI. Over the last decade, the large-scale magnetic properties of solar-mass stars were studied by several authors across the PMS. As can be seen in Fig.~\ref{fig:confusogram}, their magnetic fields evolve from strong, axisymmetric, and mostly dipolar when the stars are fully convective, to fainter and complex fields as their internal structure changes, a radiative core develops, and they go toward the MS. However, JH~223 falls in a region of the HRD containing very few subsolar mass objects, which we do not have much information on with respect to the surface magnetic field structure. Although JH~223 is fully convective, as shown by V2247 Oph and V347 Aur, the only other PMS stars with known magnetic field topologies and masses less than 0.5 M$_{\odot}$, the three systems already show complex and rather faint surface magnetic fields at an early age, in contrast to solar-type stars. On the MS, very-low-mass, fully convective, M dwarfs present a dynamo bistability, with some stars showing complex and weak magnetic fields, while others present strong dipolar fields \citep{Morin11}. It would be interesting to populate the same M dwarf region in the PMS to see if the bistability phenomenon holds for stars with the same internal structure but different ages and rotational properties.

\section{Conclusion} \label{sect:conclusions}

We analyzed photometric, spectroscopic, and spectropolarimetric data of the young low-mass accreting star JH 223, aiming to investigate the stellar magnetic field topology and the star-disk interaction. The large-scale surface magnetic field is predominantly poloidal, with a dipolar component strong enough to truncate the accretion disk near the corotation radius. The observed dipper light curves of JH 223 are associated with the presence of an inner disk warp at the base of the accretion columns, obscuring the stellar surface periodically as the star rotates. The same periodicity is seen in the modulations of radial velocity and the longitudinal magnetic field. We also associated the accretion column, traced by the maximum redshifted absorption in the H$\alpha$ and He I 1083 nm circumstellar lines, with the inner disk warp, as they both occur at similar rotational phases. The accretion process in JH 223 is dynamic, changing from an unstable to a stable regime on a timescale of a few weeks, as predicted by MHD models of the star-disk interaction. This study provides evidence that very-low-mass PMS stars with low mass accretion rates remain subject to magnetospheric accretion. Finally, our findings highlight the fact that classifications of individual stars based on photometric variability (e.g., dipper, quasiperiodic, or chaotic) only capture  a transient state. Light curves represent a snapshot of the star’s condition at a given epoch, which can evolve rapidly, reinforcing the importance of multi-epoch monitoring to understand their underlying physical processes.

\begin{acknowledgements}
TPF, BZ and SHPA acknowledge financial support from CNPq, CAPES, and FAPEMIG. JB acknowledges funding from the European Research Council under the European Union's Horizon 2020 research and innovation program (grant agreement no. 742095; SPIDI: Star-Planets-Inner Disk Interactions, https://www.spidi-eu.org). AB acknowledges support from the Deutsche Forschungsgemeinschaft (DFG, German Research Foundation) under Germany's Excellence Strategy – EXC 2094 – 390783311. AF and JAS acknowledge financial support from Large Grant INAF-2024 “Spectral Key features of Young stellar objects: Wind-Accretion LinKs Explored in the infraRed (SKYWALKER)”. We thank Dmitriy Blinov and Gina Panopoulou for providing the RoboPol data and for the discussions that enriched this work.
 
This work is based on observations obtained at the Canada–France–Hawaii Telescope (CFHT), operated by the National Research Council of Canada, the Institut National des Sciences de l’Univers of the Centre National de la Recherche Scientifique of France, and the University of Hawai‘i, including data acquired through the Gemini Remote Access to CFHT’s ESPaDOnS Spectrograph. Additional observations were obtained with the Liverpool Telescope, operated on La Palma by Liverpool John Moores University at the Observatorio del Roque de los Muchachos of the Instituto de Astrofísica de Canarias, with financial support from the UK Science and Technology Facilities Council, as well as from the Las Cumbres Observatory global telescope network, the 0.91-m telescope of the Osservatorio Astrofisico di Catania, the Crimean Astrophysical Observatory, and the Skinakas Observatory, supported by the European Social Fund (ESF) and National Resources under the Operational Programme Education and Lifelong Learning. CCP was supported by the National Research Foundation of Korea (NRF) grant funded by the Korean government (MEST) (No. 2019R1A6A1A10073437)

This work made use of the LSDpy code (\url{https://github.com/folsomcp/LSDpy}) and the ZDIpy code (\url{https://github.com/folsomcp/ZDIpy}).

\end{acknowledgements}

\bibliographystyle{aa}
\bibliography{jh223} 

\begin{appendix}

\section{Observations}\label{apObs}

\subsection{LCOGT photometry} \label{LCOGT}

Photometric observations were obtained at the Las Cumbres Observatory Global Network \citep[LCOGT,][]{Brown13} with the 1\;m Sinistro telescopes from October 16 to November 30, 2019 (LCO 2019B-019, PI L. Rebull; CLN2019B-004, PI A. Bayo). We used  SDSS/PanSTARRS $g'$, $r'$, and $i'$ filters with exposure times of 400\;s, 120\;s, and 80\;s, respectively. A total of 184 images centered on JH~223 were obtained over 47 days, yielding 46, 59, and 60 photometric measurements in the $g'$, $r'$, and $i'$ filters, respectively, after discarding some low-quality images. We used the photometric catalogs provided by LCOGT's BANZAI reduction pipeline to retrieve the instrumental magnitudes of the target as well as those of nearby comparison and control stars.  We chose JH 221, a nearby object of comparable magnitude as JH~223, as a comparison star. We checked JH 221 against XEST 07-OM-002 (2MASS J04404215+2552403), another nearby star of similar brightness. We found the two stars to be nonvariable over the observed time period and deduced photometric root-mean-squared (rms) uncertainties of 0.009, 0.014, and 0.010 mag in the $g'$, $r'$, and $i'$ filters, respectively.  The $g'$, $r'$, and $i'$ differential light curves obtained for JH~223 using JH 221 as a comparison star were subsequently flux calibrated to within an rms accuracy of 0.02\;mag using 33 contemporaneous Cousins $V$, $R_c$, and $I_c$ measurements of the same field obtained at Crimean Astronomical Observatory (CrAO).

\subsection{OACT photometry}
\label{Subsec:OACT_Photometry}

Photometric data were taken at the {M. G. Fracastoro} station (Serra La Nave, Mt. Etna, 1750\;m a.s.l.) of the {Osservatorio Astrofisico di Catania} (OACT, Italy) from October 21, 2019, to February 25, 2020, with the 0.91\,m telescope.
The CCD camera adopts a Kodak KAF1001E\footnote{\scriptsize \tt http://sln.oact.inaf.it/sln\_old/dmdocuments/ccd91rappint2-07.pdf} 1k$\times$1k chip 
that, with a focal reducer, covers a field of view (FOV) of about 11.5$\times$11.5 arcminutes.
We observed JH~223 with a set of broad band Bessel filters ($V$, $R$, $I$) as well as two narrow band filters, 
$H\alpha_9$ and $H\alpha_{18}$, centered at 656.8\,nm and at 676.4\,nm, whose full widths at half-maximum are 9 and 
18\,nm, respectively. We collected a total of 43, 70, and 73 useful images in $V$, $R$, and $I$ bands, respectively,
with exposure times of 30--40\,s, 50--80\,s, and 90--180\,s, in the respective bands. Regarding the narrow band photometry, we collected 39 images in the $H\alpha_9$ filter, with exposure times of 180--300\,s, and 78 images in the $H\alpha_{18}$ filter, with exposure times of 90--150\,s. The latter were typically acquired just before and after the H$\alpha_9$ frames to take sky transparency variations into account.

The broad band $V$, $R$, and $I$ photometry was calibrated using the following procedure. The stars in the standard area GD\,71 \citep{Landolt09} were observed in the nights with the best photometric conditions to calculate the zero points and transformation coefficients to the Johnson-Cousins system. These transformations were applied to the two non-variable stars in the FOV of JH\,223, namely XEST\,07-OM-002 and XEST\;07-OM-001 (2MASS\,J04403830+2554274), which were 
then used as comparisons to derive the magnitudes of JH~223 in each image\footnote{We derive the following magnitudes: XEST 07-OM-002: $V=16.60 \pm 0.30$, $R_c$=$15.36 \pm 0.04$, $I_c$=$13.94 \pm 0.05$; XEST\;07-OM-001 $V=16.29 \pm 0.38$, $R_c$=$14.65 \pm 0.04$, $I_c=12.95 \pm 0.04$. These values agree within 0.1--0.2 mag with those derived for the same stars at CrAO. As measurement errors are usually on the order of a few 0.01 mag, this 0.1--0.2 mag systematic uncertainty defines the accuracy of the photometric calibration reported here.}.

For details about the OACT data reduction and the narrow band H$\alpha$ photometry, the reader is referred to \citet{Frasca18}. We remark that the $CI_{\rm H\alpha}$\,=\,$H\alpha_{18}-H\alpha_9$ is basically a measure of the intensity of the H$\alpha$ emission in units of the continuum that can be converted into H$\alpha$ equivalent width ($EW_{\rm H\alpha}$), using the calibrations proposed by \citet{Frasca18}. Thanks to the contemporaneous GRACES and Keck high-resolution spectra, we have improved the aforementioned $CI_{\rm H\alpha}-EW_{\rm H\alpha}$ calibrations by means of synthetic color indices measured on the latter ones, integrating them in the passbands of the OACT $H\alpha$ filters.
   
\subsection{CrAO photometry}

Short (60, 15, and 5\;s in $V$, $R_j$, and $I_j$, respectively)  and long (180\;s, 180\;s, and 120\;s in $V$, $R_j$, and $I_j$, respectively) exposure images were obtained from September 2 to November 19, 2019, at the Crimean Astronomical Observatory through Johnson's $V$, $R_j$ and $I_j$ filters, using the AZT-11 1.25\;m telescope equipped with the CCD camera ProLine PL23042. CCD images were bias-subtracted and flatfield corrected following a standard procedure. On short exposures, we used XEST 07-OM-002 as a comparison star, for which we derived $V=16.83$, $(V_j-R_j)=2.15$, $(V_j-I_j) = 2.79$. The resulting photometric errors on long exposures typically amount to 0.09, 0.02, and 0.02\;mag in the $V$, $R_j$, and $I_j$ filters, respectively. We finally converted the magnitude and colors of JH~223 obtained in Johnson's system to Cousins's, using the following relationships: $(V_c-R_c) = (V_j-R_j) \times 0.71652 - 0.03206$; $(V_c-I_c) = (V_j-I_j) \times 0.7822  + 0.0065$ \citep{Landolt83}. We collected a total of 50 useful $V$, $R_j$, and $I_j$ images between September 02 and November 19, 2019.

\subsection{Liverpool photometry}

SDSS $u'$, $g'$, $r'$, and $i'$ photometry was performed with the 2\;m Liverpool Telescope at the Roque de los Muchachos Observatory (La Palma, Spain) using the IO:O camera equipped with the E2V detector. A total of 48 images were obtained from November 6 to 30, 2019, centered on JH~223 with a 10 arcmin square FOV, and with exposure times of 580\;s, 90\;s, 20\;s, and 10\;s in the $u'$, $g'$, $r'$, and $i'$ filters, respectively. The images were reduced by the IO:O pipeline, which includes bias subtraction, trimming of the overscan regions, and flat fielding. PSF photometry was performed with IRAF/DAOPHOT on the reduced images for JH~223, as well as for the two comparison stars XEST 07-OM-001 and XEST 07-OM-002.  We estimated the photometric errors from the differential light curve computed between the 2 comparison stars, which yielded an rms error of 0.010\;mag in $g'$, $r'$, and $i'$ filters. The objects are too faint in the $u$ band to yield reliable photometry, and measurements in this filter were thus discarded. We computed the differential light curve between JH~223 and XEST 07-OM-001, and calibrated it using contemporaneous LCOGT photometry.

\begin{table*}
    \caption{Polarisation measurements of JH~223 obtained with the RoboPol photopolarimeter.}
    \centering
    \begin{tabular}{ccccccc}
    \hline\hline
    HJD     & Pol degree & Pos. angle & $Q$ & $U$ & $E$ & Filt.\\
    (+2,457,000) & (\%) & (deg.) & (\%) & (\%) \\
    \hline
1771.580893 & 2.17 $\pm$ 0.18 & 17.8 $\pm$ 2.4 & 1.76 $\pm$ 0.19 & 1.26 $\pm$ 0.17 & 0.41 & $R_c$ \\
1771.580891 & 2.39 $\pm$ 0.28 & 22.6 $\pm$ 3.4 &  1.68 $\pm$ 0.28 & 1.70 $\pm$ 0.28 & 0.41 & SDSS $g$ \\
1787.486404 & 1.75 $\pm$ 0.16 & 26.0 $\pm$ 2.8 & 1.08 $\pm$ 0.18 & 1.38 $\pm$ 0.15 & 5.21 & $R_c$ \\
\hline
    \end{tabular}
    \label{tab:pol}
    \tablefoot{Col. 1 lists the heliocentric Julian date, Col.~2 the polarisation degree, 
Col.~3 the position angle, Cols.~4 and 5 the Stokes parameters $Q$ and $U$, 
Col.~6 the rotational cycle, and Col.~7 the filter used.}
\end{table*}

\subsection{K2 photometry}

JH~223 (EPIC 248006676) was observed by the K2 satellite during Campaign 13, which took place over 80 days from March 8 to May 23, 2017, namely, about 2.5 years before our campaign. The observations were performed in a broadband filter (420--900\;nm) with measurements taken at a cadence of 30 minutes. The K2 light curve was reduced by A. M. Cody using a 3-pixel circular aperture with background subtraction and subsequent correction of pointing jitter, as described in \citet{Cody18} and \citet{Cody22}\footnote{https://archive.stsci.edu/hlsp/k2yso}.

\subsection{TESS photometry}

The TESS satellite observed JH~223 during two epochs, from September 16 to November 05 in 2021 (sectors 43 and 44) and from September 20 to November 11 in 2023 (sectors 70 and 71), a few years after our campaign. The observations were obtained using the red-optical bandpass (600--1000\;nm) with a cadence of 2 minutes. In this work, we selected the reduction made by the Science Processing Operations Center (SPOC) pipeline \citep{2016SPIE.9913E..3EJ}.

\subsection{Skinakas observatory photopolarimetry}

Observations were performed over two nights, on October 14 and 30, 2019, with the RoboPol photopolarimeter \citep{Ramaprakash19} mounted on the 1.3\;m telescope at Skinakas Observatory, Crete,  gathering two measurements in the Cousins-$R$ band and an additional one in the SDSS $g$ filter. The target was placed in the center of the field of view, where a focal-plane mask produces shadows to reduce the background light. Polarized and unpolarized standard stars were observed on the nights of the target observations and were used in the calibration. The data were reduced using the RoboPol pipeline \citep{REBOPOL-PIPELINE-2014}, which performs differential aperture photometry on the set of four target images produced by the instrument to measure the relative Stokes parameters $Q$ and $U$. The results are listed in Table~\ref{tab:pol}.

\subsection{Gemini/GRACES optical spectroscopy}\label{GRACES}

The optical spectroscopic observations were obtained with the Gemini Remote Access to CFHT ESPaDOnS Spectrograph (GRACES), which consists of the ESPaDOnS high-resolution spectrograph \citep{Donati03} at the Canada-France-Hawaii Telescope (CFHT) connected through two $270$\;m optical fibers to the 8.1\;m Gemini North telescope \citep{chene2014graces}. The observations were obtained over 4 nights between November 7, 2019, and January 20, 2020, with a resolution $R \sim 40000$ in the wavelength range of $370$\;nm to $950$\;nm (proposal ID: GN-2019B-Q-101;\; PI: S. Alencar). The data reduction was done with the \verb|OPERA| pipeline \citep{martioli2012open}, which includes bias subtraction, flat-field correction, and wavelength calibration. All spectra were continuum normalized.

\subsection{Keck I optical spectroscopy}\label{Keck}

We also obtained optical spectroscopic observations with the Keck I Telescope using the High-Resolution Echelle Spectrometer \citep[HIRES,][]{vogt-HIRES}, with a spectral resolution of $R \sim 48000$, covering the wavelength range from 470\;nm to 930\;nm. A total of four spectra were acquired between November 29 and December 1, 2019. However, only three were included in the analysis, as one exhibited a low S/N. The data reduction was done using the \verb|MAKEE|\footnote{\url{https://astro.caltech.edu/~tb/makee/}} reduction pipeline written by T. Barlow. 

\subsection{CFHT/SPIRou near-infrared spectropolarimetry} \label{SPIRou}

JH~223 was observed for eight non-consecutive nights in November 2019 using the near-IR spectropolarimeter and velocimeter SPIRou at the CFHT (proposal ID: 19BF02; PI.: J. Bouvier). SPIRou is a stabilized high-resolution near-infrared (nIR) spectropolarimeter that covers wavelengths between $950$\;nm and $2350$\;nm with a resolution $R \sim 70000$ \citep{10.1093/mnras/staa2569}. SPIRou observations consist of sequences of four subexposures taken at distinct orientations of the quarter-wave Fresnel rhombs, where each subexposure (integrated over $903$\;s for JH~223) includes two spectra with orthogonal polarization states \citep{1997MNRAS.291..658D,2009PASP..121..993B}. 

The data reduction was performed using the \verb|APERO| pipeline software \citep{2022PASP..134k4509C}. In addition to the standard extraction, flat-field correction, and wavelength calibration, \verb|APERO| performs the polarimetric data-reduction, combines subexposures to extract normalized unpolarized (Stokes $I$), circularly polarized (Stokes $V$), and null (Stokes $N$) spectra that are telluric-corrected following the method described in \citet{ACC22}. The SPIRou data are publicly available from the Canadian Astronomy Data Center\footnote{\url{https://www.cadc-ccda.hia-iha.nrc-cnrc.gc.ca/en/cfht/}}.

\section{Circumstellar lines from GRACES, Keck, and SPIRou observations}\label{apCirLines}

\begin{figure*}[ht!]
    \centering \includegraphics[width=\linewidth]{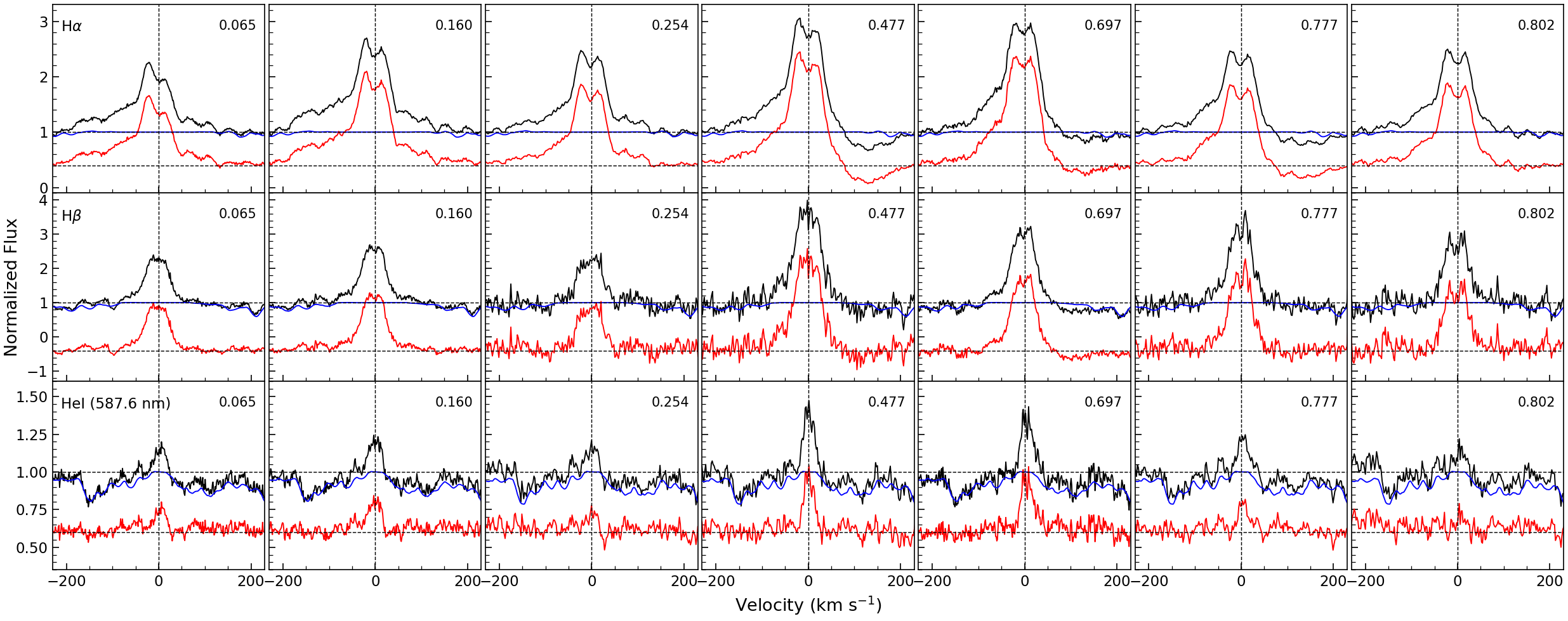}
    \caption{Circumstellar emission lines of $\mathrm{H{\alpha}}$ (top row), $\mathrm{H{\beta}}$ (middle row), and He I 587.6\;nm (bottom row) from the GRACES (Cols. 3, 4, 6, and 7) and Keck (Cols. 1, 2, and 5) observations. The blue curves show the spectrum of TWA 7, used as a photospheric template. The black and red curves correspond to the spectrum of JH~223 before and after removal of the photospheric contribution, respectively. The photosphere-subtracted spectra are vertically shifted for clarity.}
   \label{fig:residualGRACES}
\end{figure*}

\begin{figure*}[ht!]
    \centering
   \includegraphics[width=\linewidth]{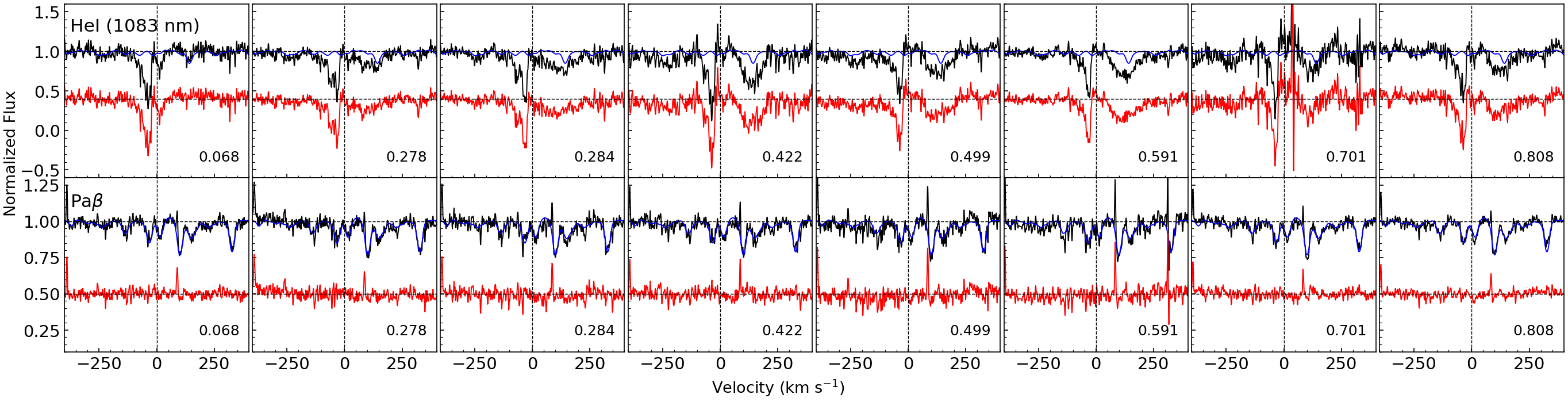}
    \caption{Similar to Fig.~\ref{fig:residualGRACES}, but for the He I 1083\;nm (top) and Pa$\beta$ (bottom) emission lines from the SPIRou observations.}
   \label{fig:residuals-SPIRou}
\end{figure*}

Figure \ref{fig:residualGRACES} shows the $\mathrm{H{\alpha}}$, $\mathrm{H{\beta}}$, and He I 587.6 nm emission lines from GRACES and Keck observations before (black) and after (red) the removal of the photospheric contribution. For the near-infrared circumstellar emission lines of He I 1083\;nm and Pa$\beta$, the lines profiles obtained from SPIRou observations are shown in Figure~\ref{fig:residuals-SPIRou}.

\begin{figure}[ht]

   \centering   \includegraphics[width=0.9\linewidth]{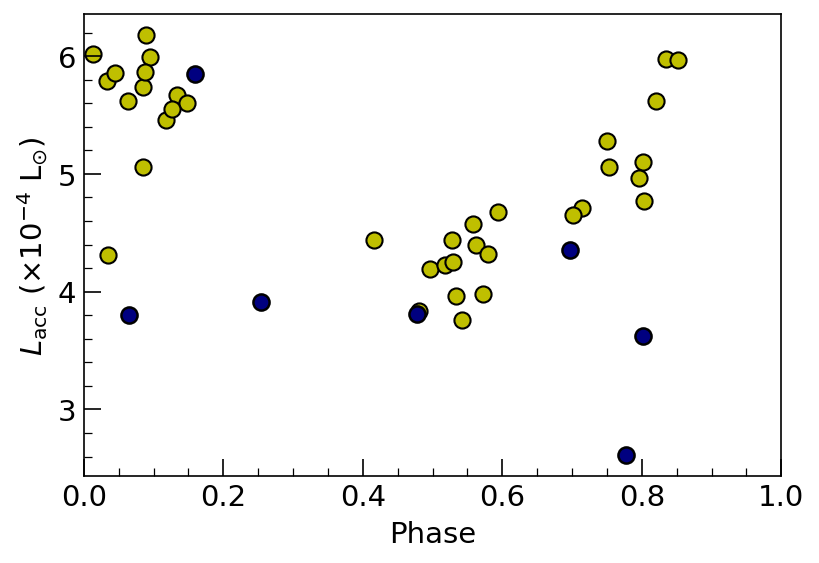}
    \caption{Accretion luminosity of $\mathrm{H{\alpha}}$ as a function of rotational phase. The yellow circles represent the accretion luminosities obtained from the equivalent widths derived from the narrowband color-index filters $CI_{\rm H\alpha}$, while the blue circles correspond to the values derived from the GRACES and Keck spectra.}
              \label{fig:lacc_vs_phase}
\end{figure}

Table~\ref{tab:parhalpha} summarizes the equivalent widths, line fluxes, and line luminosities of the $\mathrm{H{\alpha}}$ and $\mathrm{H{\beta}}$ lines measured from the GRACES and Keck spectra. As described in Sect.~\ref{sec: Mass accretion rate}, these quantities were used to derive the accretion luminosity and mass accretion rate of JH~223, which are also reported in Table~\ref{tab:parhalpha}. In Fig.\ref{fig:lacc_vs_phase}, we illustrate the rotational modulation found in the accretion luminosity derived from the $\mathrm{H{\alpha}}$ line.

\begin{table*}
\caption{Accretion parameters calculated with the $\mathrm{H{\alpha}}$ (top) and $\mathrm{H{\beta}}$ (bottom) lines from optical spectra (GRACES and Keck).}
\centering
\begin{tabular}{cccccccc}
\hline
\hline
 $E$ & $EW_{\mathrm{H\alpha}}$ &   $m_{R}$  & $F_\mathrm{H{\alpha}}$ & $L_\mathrm{H{\alpha}}$ & $L_{acc}$ & $\dot{M}_{acc}$ \\ 
&    (\AA)     &           & (erg\,s$^{-1}$\,cm$^{-2}$)  & (erg\,cm$^{-2}$)  & (L$_{\odot}$)  & (M$_{\odot}$/yr) \\ \hline
 7.477 &  3.93  &  14.333  &  $4.28 \times 10^{-14}$   & $1.04 \times 10^{29}$  &  $3.81 \times 10^{-4}$   & $4.34 \times 10^{-11}$    \\ 
  7.777 &  2.66  &  14.276 &  $3.06 \times 10^{-14}$   &  $0.75 \times 10^{29}$ &  $2.61 \times 10^{-4}$   & $2.97 \times 10^{-11}$ \\ 
  9.254 &   3.38  & 14.142 &  $4.38 \times 10^{-14}$  &  $1.07 \times 10^{29}$ &  $3.91 \times 10^{-4}$   & $4.46 \times 10^{-11}$ \\ 
  14.065&   3.07  & 14.068 &  $4.27 \times 10^{-14}$  &  $1.04 \times 10^{29}$ &  $3.80 \times 10^{-4}$   & $4.33 \times 10^{-11}$ \\ 
  14.160 &   4.56  & 14.081 &  $6.26 \times 10^{-14}$  &  $1.53 \times 10^{29}$ &  $5.86 \times 10^{-4}$   & $6.67 \times 10^{-11}$ \\ 
  14.697 &  4.07  & 14.243 &  $4.81 \times 10^{-14}$  &  $1.17 \times 10^{29}$ &  $4.35 \times 10^{-4}$   & $4.95 \times 10^{-11}$ \\ 
  29.802 &   3.18  & 14.148 &  $4.10 \times 10^{-14}$ & $1.00 \times 10^{29}$  & $3.62 \times 10^{-4}$   &   $4.13 \times 10^{-11}$   \\ 
\hline

\end{tabular}
  \label{tab:parhalpha}
\end{table*}

\begin{table*}
\centering
\begin{tabular}{ccccccc}
\hline
\hline
$E$ &  $EW_{\mathrm{H\beta}}$ &  $m_{V}$  &  $F_\mathrm{H{\beta}}$ & $L_\mathrm{H{\beta}}$ & $L_{acc}$ & $\dot{M}_{acc}$ \\ 
&   (\AA)   &       & (erg\,s$^{-1}$\,cm$^{-2}$)  & (erg\,cm$^{-2}$)  &  (L$_{\odot}$)   & (M$_{\odot}$/yr) \\ \hline
 7.477 &  3.23  &  15.847 &  $1.62\times 10^{-14}$   &  $3.97\times 10^{28}$ & $8.02\times 10^{-4}$   & $9.15 \times 10^{-11}$\\ 
  7.777  &  2.46  &  15.744 &   $1.36 \times 10^{-14}$   &  $3.33 \times 10^{28}$ &  $6.56 \times 10^{-4}$   & $7.48 \times 10^{-11}$  \\ 
 9.254  &  1.42  & 15.589  &   $0.91 \times 10^{-14}$   & $2.22 \times 10^{28}$  &  $4.14 \times 10^{-4}$   & $4.72 \times 10^{-11}$ \\ 
  14.065&   2.07  & 15.527 &  $1.40 \times 10^{-14}$  &  $3.41 \times 10^{28}$ &  $6.76 \times 10^{-4}$   & $7.70\times 10^{-11}$ \\ 
  14.160 &   2.66  & 15.516 &  $1.81 \times 10^{-14}$  &  $4.44 \times 10^{28}$ &  $9.11 \times 10^{-4}$   & $10.4 \times 10^{-11}$ \\ 
  14.697 &  2.64  & 15.720 &  $1.49 \times 10^{-14}$  &  $3.65 \times 10^{28}$ &  $7.30 \times 10^{-4}$   & $8.32 \times 10^{-11}$ \\ 
  29.802  &  2.18  &  15.588 &   $1.39 \times 10^{-14}$  & $3.41 \times 10^{28}$  &  $6.75 \times 10^{-4}$  &  $7.69 \times 10^{-11}$   \\
\hline
\end{tabular}
\tablefoot{Column 1 provides the rotational cycle ($E$), Col.~2 the $EW$ of lines, and Col.~3 the corresponding magnitude of each observation. Columns 4, 5, 6, and 7 represent, respectively, the flux of the $\mathrm{H{\alpha}}$ ($\mathrm{H{\beta}}$) line, the luminosity of the $\mathrm{H{\alpha}}$ ($\mathrm{H{\beta}}$) line, the accretion luminosity and the mass accretion rate.}
\end{table*}

\section{Details about the ZDI reconstruction}
\subsection{The $\chi^{2}$ aiming criteria} \label{apChi}
\begin{figure}
    \centering
\includegraphics[width=0.5\textwidth]{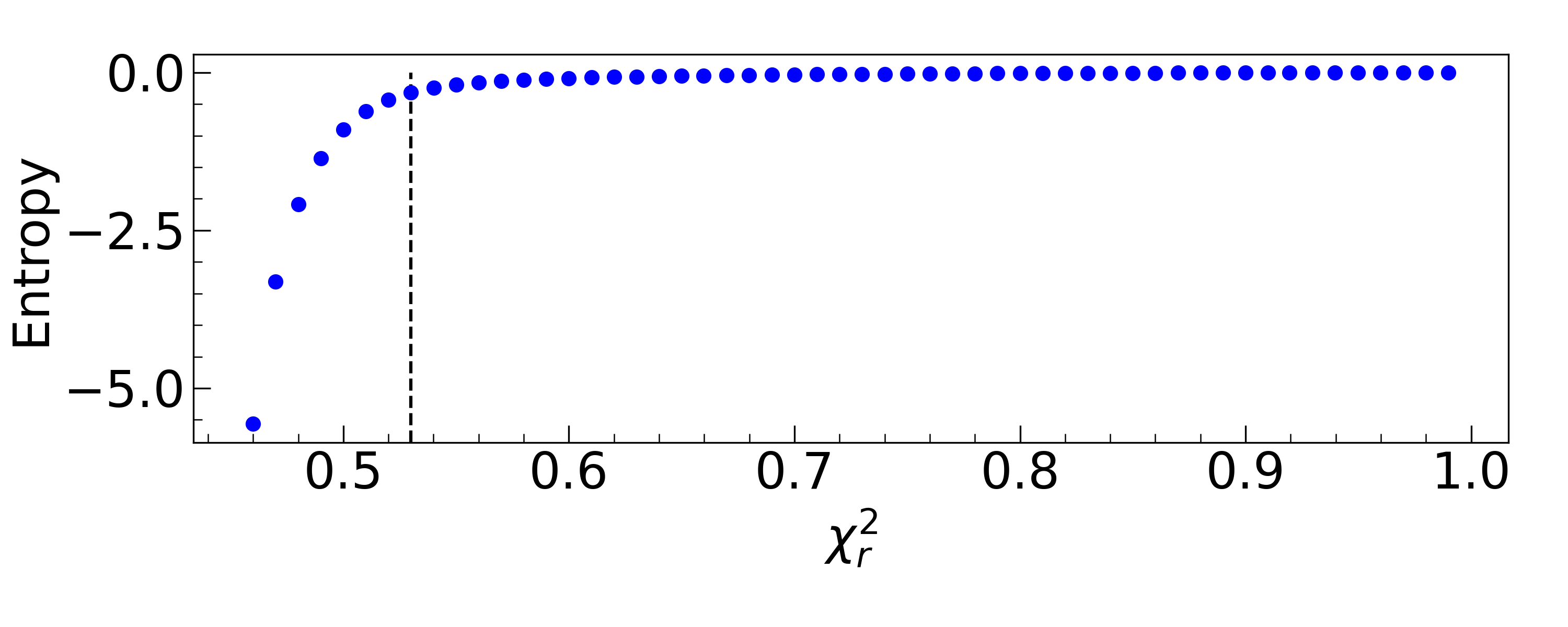}
    \caption{Entropy versus $\chi^{2}_{r}$ for several models of the reconstructed brightness images of JH~223. Each point corresponds to a converged ZDI solution. The vertical line indicates the inflection point of $\chi^{2}_{r,sf} = 0.53$ that was determined using the Kneedle algorithm.}
    \label{fig:scaling}
\end{figure}

We computed a series of ZDI reconstructions aiming at different values of $\chi^{2}_{r}$. We looked at the variation of the image entropy as a function of $\chi^{2}_{r}$ and applied the Kneedle algorithm \citep{satopaa2011finding}, implemented in the Python package \verb|Kneed|\footnote{\url{https://pypi.org/project/kneed/}}, as an approach to search for $\chi^{2}_{r,sf}$. The \verb|Kneed| algorithm identifies inflection points in curves characterized by negative or positive concavity in discrete datasets, using the mathematical definition of curvature for continuous functions. According to \citet{satopaa2011finding}, this method has shown greater effectiveness compared to other numerical approaches proposed to detect knees in discrete data, such as Angle-based and Menger Curvature \citep{basseville1993detection,bollinger2002bollinger,zhao2008knee}. Figure~\ref{fig:scaling} shows an example of the Kneedle algorithm applied to the reconstructed brightness images of JH~223. Through the \verb|Kneed| code, we obtained the inflection point $\chi^{2}_{r,sf} = 0.53$. We applied the correction to the error bars of all LSD profiles, which allowed us to target a unit $\chi^{2}_{r}$ in the following image reconstructions.

\subsection{Best entropy model}
\begin{figure}
    \centering
    \includegraphics[width=\linewidth]{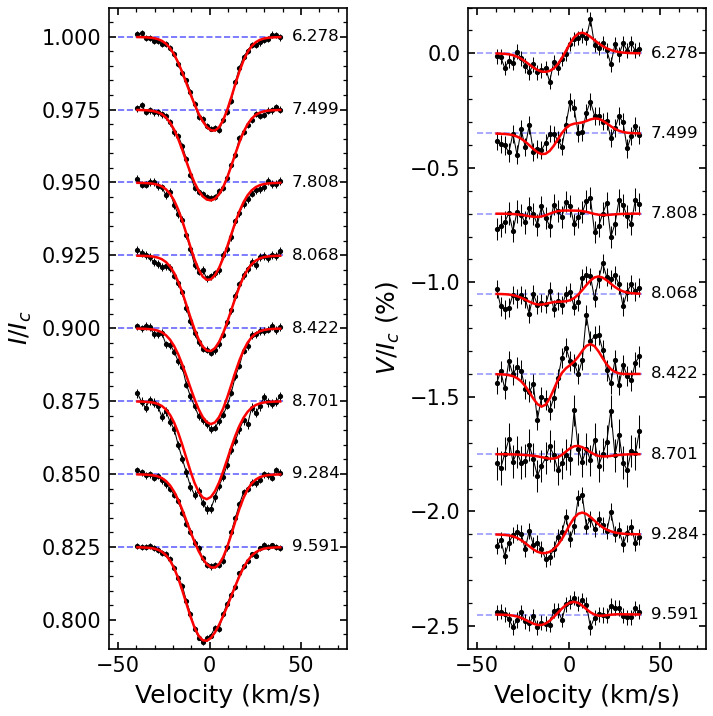}
    \caption{ZDI fit of the Stokes $I$ (left) and $V$ (right) LSD profiles of JH~223. Observed (black) and maximum-entropy synthetic (red) profiles are shown. Dashed blue lines indicate the continua. Profiles are vertically shifted for clarity and labeled by rotational phase.}
    \label{fig:stokesIV}
\end{figure}

Figure~\ref{fig:stokesIV} shows the maximum-entropy solution obtained by applying ZDI technique to the time-series of Stokes $I$ and $V$ LSD profiles computed from the SPIRou dataset.

\end{appendix}

\end{document}